\documentclass[accepted=2021-06-15,letterpaper,aps,twocolumn,notitlepage,amsmath,amstex,amssymb,citeautoscript,longbibliography,bibnotes]{quantumarticle}
\pdfoutput=1
\usepackage{natbib}
\usepackage[english]{babel}
\usepackage{letltxmacro}
\usepackage{latexsym}
\LetLtxMacro{\ORIGselectlanguage}{\selectlanguage}
\makeatletter
\DeclareRobustCommand{\selectlanguage}[1]{%
  \@ifundefined{alias@\string#1}
    {\ORIGselectlanguage{#1}}
    {\begingroup\edef\x{\endgroup
       \noexpand\ORIGselectlanguage{\@nameuse{alias@#1}}}\x}%
}
\newcommand{\definelanguagealias}[2]{%
  \@namedef{alias@#1}{#2}%
}
\makeatother
\definelanguagealias{en}{english}
\definelanguagealias{English}{english}

\usepackage{graphicx}
\usepackage{bm}
\usepackage{algorithm,algpseudocode}
\usepackage{physics}
\usepackage{hyperref}
\hypersetup{
    bookmarks=false,         
    unicode=false,          
    pdftoolbar=false,        
    pdfmenubar=true,        
    pdffitwindow=false,     
    pdfstartview={FitH},    
    pdftitle={Quantum Annealing Initialization of the Quantum Approximate Optimization Algorithm},    
    pdfauthor={Stefan H. Sack, Maksym Serbyn},     
    pdfsubject={QA,QAOA},   
    pdfcreator={},   
    pdfproducer={}, 
    pdfkeywords={QA,QAOA}, 
    pdfnewwindow=true,      
    colorlinks=true,       
    linkcolor=black,          
    citecolor=blue,        
    filecolor=magenta,      
    urlcolor=blue           
}
\usepackage{hyperref}
\usepackage{tikz}
\newcommand{\slope}{{\delta t}}
\newcommand{\im}{{\rm i}}
\begin{document}
\title{Quantum Annealing Initialization of the Quantum Approximate Optimization Algorithm}
\author{Stefan H. Sack}
\orcid{0000-0001-5400-8508}
\email{stefan.sack@ist.ac.at}
\author{Maksym Serbyn }
\orcid{0000-0002-2399-5827}
\email{maksym.serbyn@ist.ac.at}
\affiliation{IST Austria, Am Campus 1, 3400 Klosterneuburg, Austria
}
\date{June 28, 2021}
\begin{abstract}
The quantum approximate optimization algorithm (QAOA) is a prospective near-term quantum algorithm due to its modest circuit depth and promising benchmarks. However, an external parameter optimization required in the QAOA could become a performance bottleneck. This motivates studies of the optimization landscape and search for heuristic ways of parameter initialization. In this work we visualize the optimization landscape of the QAOA applied to the MaxCut problem on random graphs, demonstrating that random initialization of the QAOA is prone to converging to local minima with sub-optimal performance. We introduce the initialization of QAOA parameters based on the Trotterized quantum annealing (TQA) protocol, parameterized by the Trotter time step. We find that the TQA initialization allows to circumvent the issue of false minima for a broad range of time steps, yielding the same performance as the best result out of an exponentially scaling number of random initializations. Moreover, we demonstrate that the optimal value of the time step coincides with the point of proliferation of Trotter errors in quantum annealing. Our results suggest practical ways of initializing QAOA protocols on near-term quantum devices and reveal new connections between QAOA and quantum annealing. 
\end{abstract}

\maketitle

\section{Introduction}
Recent technological advances have led to a large number of implementations~\cite{arute2020quantum, arute2020hartree, arute2020observation, wright2019benchmarking} of so-called Noisy Intermediate-Scale Quantum (NISQ) devices~\cite{preskill2018quantum}. These machines, which allow to manipulate a small number of imperfect qubits with limited coherence time, inspired the search for practical quantum algorithms. The quantum approximate optimization algorithm (QAOA)~\cite{farhi2014quantum} has emerged as a promising candidate for such NISQ devices~\cite{zhou2018quantum, crooks2018performance, willsch2020benchmarking}. 

\begin{figure}[t]
    \centering
    \includegraphics[width=\columnwidth]{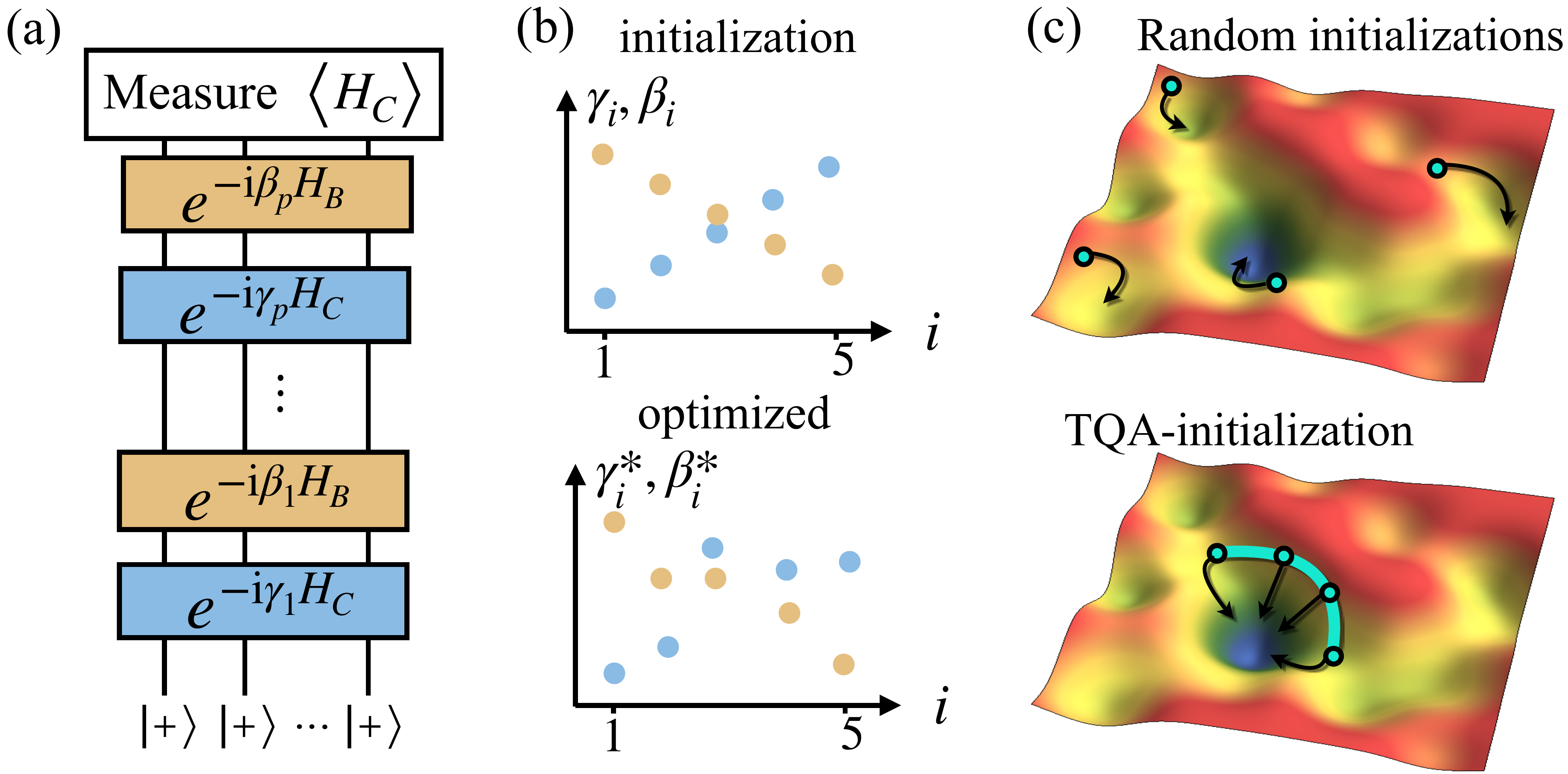}
    \caption{(a) The circuit that prepares a quantum state in the QAOA is parametrized by a set of $2p$ angles $\gamma_i,\beta_i$. For the MaxCut problem, that is considered in the main text, the unitaries can be expressed using single and two qubit gates that are readily available on current NISQ devices. (b)~The optimization of $\langle{H_C}\rangle$ is launched from a certain guess of parameters, state preparation and measurements are iterated until the algorithm converges to a set of optimized angles $\gamma^*_i,\beta^*_i$. (c) The cartoon of the cost function $\langle{H_C}\rangle$ landscape as a function of variational parameters shows that random initializations are prone to converge to sub-optimal local minima. In contrast, the family of TQA initializations proposed in this work converges to the (nearly) optimal minimum.
      }
    \label{fig1}
\end{figure}

The QAOA is a variational hybrid quantum algorithm where the classical computer operates a NISQ device. The computer is responsible for the optimization of the cost function over a set of variational parameters. The cost function is calculated using a NISQ device that prepares a quantum state corresponding to chosen parameters and performs quantum measurements. In QAOA of depth $p$ the wave function is prepared by a unitary circuit parametrized by $2p$ parameters, see Fig.~\ref{fig1}(a). Each of the $p$ layers consist of two unitaries: the first is generated by a classical Hamiltonian $H_C$ that encodes the cost function of a combinatorial optimization problem, and the second is generated by the mixing quantum Hamiltonian, $H_B$.

While the $p=1$ limit of QAOA allows for analytic considerations and derivation of performance guarantees~\cite{farhi2014quantum}, subsequent work suggested that higher depth $p$ may be required in order to achieve a quantum advantage~\cite{bravyi2019obstacles, crooks2018performance}. However, increasing $p$ leads to a progressively more complex optimization landscape, that is characterized by a large number of local suboptimal minima~\cite{zhou2018quantum, willsch2020benchmarking, guerreschi2019qaoa, shaydulin2019multistart}, see Fig.~\ref{fig1}(c). The convergence of classical optimization algorithms into such sub-optimal solutions was demonstrated to be a potential bottleneck of QAOA performance as finding a nearly optimal minimum usually requires exponential in $p$ number of initializations of the classical optimization algorithm~\cite{farhi2014quantum, zhou2018quantum}.  Note, that the problem of sub-optimal local minima is different from that of barren plateaus~\cite{mcClean2021barren, holmes2021connecting}, i.e. large regions in parameter space with vanishing gradients, since barren plateaus are associated with circuit depths $p$ polynomial in system size $N$~\cite{cerzo2021cost}, beyond what is typically considered in the QAOA. 

The complexity of the energy landscape of large-$p$ QAOA has motivated the search for heuristic ways of improving the convergence to a (nearly) optimal minimum values of the variational parameters. Recent work has demonstrated a concentration of the QAOA landscape for typical problem instances~\cite{Brandao}, which implies the existence of a typical landscape and hints at the fact that the same variational parameter choice may work between different problem instances or sizes.  A particular example of such a heuristic was proposed in Ref.~\cite{zhou2018quantum} which constructs a good initialization for the QAOA at level $p+1$ using the solution at level $p$, thus requiring a polynomial in $p$ number of optimization runs. Other approaches, such as reusing parameters from similar graphs~\cite{shaydulin2019multistart}, using an initial state that encodes the solution of a relaxed problem~\cite{egger2020warm}, or utilizing machine learning techniques to predict QAOA parameters~\cite{alam2020accelerating, khairy2019learning} were also proposed.

In this work we propose a different approach to the QAOA initialization, based on the relation between QAOA and the quantum annealing algorithm. Quantum annealing uses adiabatic time evolution to find the lowest energy state of $H_C$, but often requires unfeasible evolution time $T$~\cite{albash2018adiabatic}. We explore the observation that Trotterization of unitary evolution in quantum annealing provides a particular choice of parameters for the QAOA~\cite{farhi2014quantum}. This leads us to introduce a one-parameter family of Trotterized quantum annealing (TQA) initializations for QAOA, controlled by the time step or, equivalently, total time used in adiabatic evolution.

The central result of our work is the demonstration that TQA initialization for QAOA gives comparable performance to the search over an exponentially scaling number of random initializations. To this end, we establish that TQA initialization leads to convergence of the QAOA to a nearly optimal minimum for a certain range of time steps, see Fig.~\ref{fig1}(c) for visualization. Furthermore, we identify the optimal time step of the TQA initialization and suggest a purely experimental way of fixing this parameter. 

Our work reveals a connection between intermediate-$p$ QAOA and short-time quantum annealing. Previous studies~\cite{farhi2014quantum,crooks2018performance,zhou2018quantum} established a correspondence between quantum annealing with long annealing times and the QAOA protocol with large $p$ (potentially increasing exponentially with the problem size). More recent work proposed quantum annealing inspired initialization strategies for the so-called `bang-bang' modification of the QAOA~\cite{liang2020investigating} that however also corresponds to large circuit depths. Our work is different from this context, since we establish that the best performance is achieved for a very \emph{coarse discretization} of quantum annealing, resulting in a realistic circuit depth. We show the existence of an optimal step for TQA discretization that does not depend on problem size and QAOA depth. This suggests an intimate relation between QAOA and TQA, since the optimal value of the time step is in close correspondence to the point where proliferation of Trotter error occurs in TQA~\cite{heyl2019quantum}.

The remainder of the paper is organized as follows: in Section~\ref{optimization_landscape} we introduce the QAOA, visualize its optimization landscape and show that most random initializations concentrate around sub-optimal local minima. Next, in Section~\ref{tqa_init} we discuss TQA and the corresponding initialization and show that it avoids converging at sub-optimal local optima. Finally, in Section~\ref{discussion} we summarize the results, discuss its implications and potential future work.

\section{Optimization landscape of QAOA}\label{optimization_landscape}
\subsection{QAOA for MaxCut problems}

As we discussed in the Introduction, the QAOA  is often applied to hard combinatorial optimization problems.  In what follows we concentrate on the problem of finding a maximal cut (MaxCut) in a given graph which has become one of standard tasks used to benchmark the QAOA~\cite{willsch2020benchmarking, zhou2018quantum}. Finding the maximum cut is an ${NP}$-hard combinatorial optimization problem, though efficient classical algorithms exist that yield good approximate solutions. Notably, the Goemans-Williamson algorithm yields a cut that is at least $88\%$ of the size of
the maximum cut in polynomial time~\cite{goemans1995improved}.  

Given a graph $G=(V, E)$ with vertices $V={1, 2, ..., N}$ and edges $E=\{\langle i,j \rangle \}$, the \emph{maximal cut} is defined as the partition that splits the vertices into two groups, maximizing the number of edges that connect vertices from different groups. Mathematically, such partition amounts to finding the global minimum of a cost function,  $C(\vec{z}) = \sum_{\langle i,j\rangle\epsilon E} z_i z_j$, where the binary variables $z_i$ correspond to the vertices of the graph, and their value $z_i = \pm 1$ encodes which partition the given vertex $i$ belongs to. The cost function $C(\vec{z})$ can be mapped into a classical spin Hamiltonian by promoting the binary variable $z_i$ to the quantum spin-1/2 operator $\sigma^z_i$. The resulting Hamiltonian,
\begin{equation}  \label{H_C}
    H_C = \sum_{\langle i,j\rangle \epsilon E} \sigma^z_i \sigma^z_j,
\end{equation}
operates on $N$ spins that reside on the vertices $V$ of the corresponding graph and interact with each other when connected by an edge. 

The QAOA uses a NISQ device to prepare the following quantum state [see Fig.~\ref{fig1}(a)]:
\begin{equation}\label{qaoa}
    |\vec{\gamma},\vec{\beta}\rangle = \prod_{i=1}^{p} e^{-\im \beta_i H_B} e^{-\im \gamma_i H_C} \ket{0}_B,
\end{equation}
where $H_C$ is classical Hamiltonian introduced above, and $H_B = -\sum_i^N \sigma^x_i$ the mixing Hamiltonian, as proposed by \textcite{farhi2014quantum}. Both operators operate on the Hilbert space corresponding to $N$ spins or, equivalently, qubits, and the initial state $\ket{0}_B= \ket{+}^{\otimes N}$  corresponds to all qubits pointing along $x$-direction, thus yielding the ground state of $H_B$. The variational parameters are obtained by minimizing the expectation value $\langle{H_C}\rangle_{\vec{\gamma},\vec{\beta}}$ as:
\begin{equation} \label{optimize}
    (\vec{\gamma^*},\vec{\beta^*}) = \arg \min_{(\vec{\gamma},\vec{\beta})} \langle {\vec{\gamma},\vec{\beta}} |{H_C}|{\vec{\gamma},\vec{\beta}}\rangle,
\end{equation}
which is typically carried out with numerical optimization routines.
To benchmark the QAOA it is useful to define the approximation ratio,
\begin{equation}\label{Eq:rdef}
r_{\vec \gamma,\vec\beta}=\frac{\langle \vec\gamma,\vec\beta|H_C| \vec\gamma,\vec\beta\rangle}{C_{\min}},
\end{equation}
which quantifies how close the expectation value of the classical Hamiltonian over the QAOA wave function is to the ground state energy of $H_C$, denoted as $C_{\min}$. For QAOA at depth $p=1$ the algorithm is guaranteed to find a cut that is at least $69\%$ the size of the optimal cut~\cite{farhi2014quantum}, while for $p>1$ analytic results are limited~\cite{wurtz2020bounds}.      

The performance of the QAOA is typically investigated over an ensemble of graphs rather than an individual realization. Below we focus on the ensemble of random 3-regular graphs, where each vertex is connected to three other vertices chosen at random. However, in the Appendix we also consider weighted 3-regular graphs and Erd\H{o}s-R\'enyi graph ensembles in order to illustrate the general applicability of our results.

\subsection{Visualizing optimization landscape}

The performance of the classical optimization in Eq.~(\ref{optimize}) strongly depends on the properties of the optimization landscape. While this landscape can be readily visualized for $p=1$, the dependence of approximation ratio $r_{\vec{\gamma}, \vec{\beta}}$ on $2p$ angles parametrizing QAOA was suggested to become progressively more complex for larger values of $p$. In order to visualize the properties of this high-dimensional landscape,  we focus below on points where  $1-r_{\vec{\gamma}, \vec{\beta}}$ achieves (local) minima. 

We quantify properties of minima using two different characteristics. First, we measure the difference between the approximation ratio of the given minimum characterized by angles $\vec \gamma,\vec\beta$ and the global minimum characterized by angles $\vec \gamma^*,\vec\beta^*$,   $ \Delta r_{\vec \gamma, \vec \beta} =r_{\vec \gamma^*,\vec\beta^*}-r_{\vec \gamma,\vec\beta}$. This definition implies that the smallest possible value of $ \Delta r_{\vec \gamma, \vec \beta}$ is $0$, and larger values of $ \Delta r_{\vec \gamma, \vec \beta}$ corresponds to local minima with poor performance (i.e.\ much larger value of cost function) compared to the global minimum. The second characteristic measures the distance between minima in parameter space, 
\begin{equation}\label{angular_distance}
    d_{\vec \gamma,\vec \beta} = \sum_{i=1}^p  \left(|\beta_i - \beta^*_i|_{\frac{\pi}{2}} + |\gamma_i - {\gamma^*_i}|_{\pi}\right),
\end{equation}
where $|\ldots|_{\alpha}$ denotes the absolute value modulo $\alpha$ which takes into account symmetries, see Appendix~\ref{app1}.

We calculate values of $ \Delta r_{\vec \gamma, \vec \beta}$ and $d_{\vec \gamma,\vec \beta}$ numerically. For a given graph realization we use $2^p$ different random initializations of variational parameters $\vec{\gamma},\vec\beta$ and optimize them using the  iterative BFGS algorithm~\cite{bfgs1, bfgs2, bfgs3, bfgs4}. The algorithm is accessed via the \texttt{scipy.optimize} Python module with default parameters~\cite{scipy}. Convergence is achieved when the norm of the gradient is less than $10^{-5}$, maximum number of iterations is set to $400 p$, where $p$ is the QAOA depth. In our simulations the routine typically converged before using up the maximum number of allowed iterations. We use the converged angles with the lowest value of $1-r_{\vec\gamma,\vec\beta}$ as an estimate for the global minimum~$\gamma^*_i,\beta^*_i$. 

\begin{figure}[h]
    \centering
    \includegraphics[width=\columnwidth]{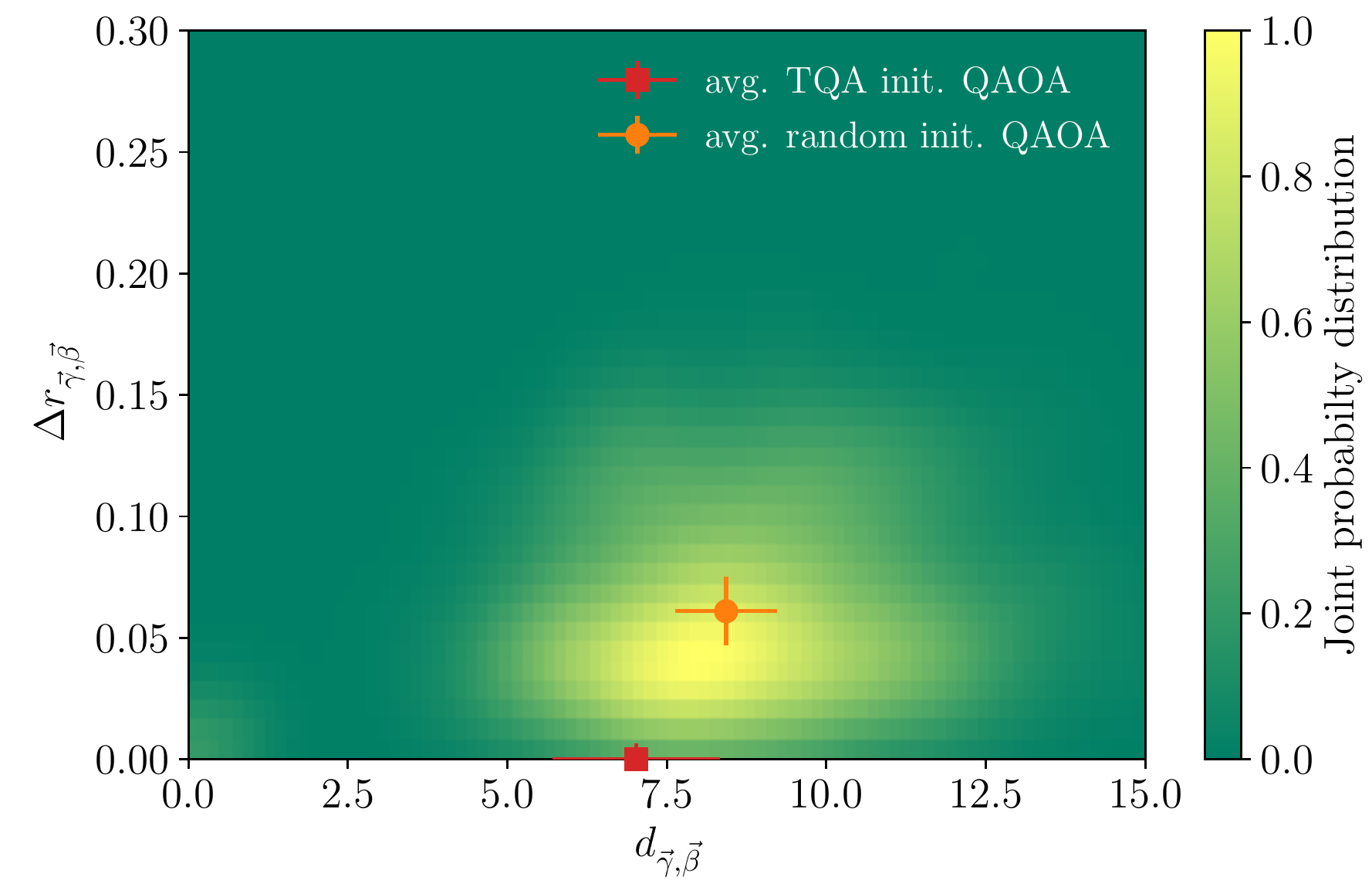}
    \caption{Joint probability distribution of distance to the global minimum in parameter space $d_{\vec\gamma,\vec\beta}$ and in terms of approximation ratio $\Delta r_{\vec\gamma,\vec\beta}$ reveals that the most probable outcome of random initialization is a convergence to sub-optimal local minima (yellow region). The orange dot corresponds to average values of $d_{\vec\gamma,\vec\beta},\Delta r_{\vec\gamma,\vec\beta}$ for random initialization.  In contrast, TQA initialization leads to a local minimum with a better approximation ratio that occasionally outperforms the best random initialization (red square, shifted from slightly negative values to $\Delta r_{\vec\gamma,\vec\beta}=0$ for improved visibility). The data is averaged over 50 random unweighted 3-regular graphs with $N=12$ vertices and QAOA at level $p=5$.}
    \label{fig2}
\end{figure}

Figure~\ref{fig2} visualizes the structure of local minima via the joint probability distribution of $ \Delta r_{\vec \gamma, \vec \beta}$ and $d_{\vec \gamma,\vec \beta}$ for 50 different graphs using Kernel Density Estimation~\cite{rosenblatt1956, parzen1962}. We observe that for QAOA with $p=5$ the most typical local minima reached from random initialization are far away from the best minimum (corresponding to $\Delta r_{\vec \gamma^*, \vec \beta^*}=0$ and $d_{\vec \gamma^*,\vec \beta^*}=0)$ both in terms of quality of approximation ratio and parameter values. While this figure illustrates a particular choice of system size and QAOA depth, a similar trend is observed for different $N$, $p$, and other graph ensembles, see Appendix~\ref{app1}. 

The tendency of random initialization to converge to suboptimal solutions highlights the importance of better initialization methods.  In the next section we investigate a family of initializations inspired by quantum annealing and demonstrate that it achieves a good approximation ratio with a suitable choice of parameters. 

\section{Trotterized quantum annealing as initialization}\label{tqa_init}

\subsection{Optimal time for TQA}
    Quantum annealing~\cite{kadowaki1998quantum, brooke1999quantum} was among the first algorithms proposed for quantum computing~\cite{farhi2001quantum, farhi2000quantum}, and was demonstrated to be universal for $T\to \infty$ and equivalent to digital quantum computing~\cite{aharonov2008adiabatic}. The general idea of quantum annealing is to prepare the ground state $\ket{0}_C$ of a classical Hamiltonian $H_C$ starting from the ground state $\ket{0}_B$ of the mixing Hamiltonian $H_B$ using adiabatic time evolution under the Hamiltonian $H(t)  =   (1-t/T) H_B + (t/T)H_C$. Practical execution of quantum annealing on NISQ devices requires discretization to represent such unitary evolution via a sequence of gates, resulting in the TQA algorithm. The first order Suzuki-Trotter decomposition allows to approximate the time evolution with $H(t)$ over time interval  $\Delta t$ as  $e^{-\im \Delta t H(t)}\approx e^{-\im \beta H_B} e^{-\im \gamma H_C}+ \mathcal{O}({\Delta t}^2)$ with $\beta=(1-{t}/{T}) \Delta t$ and $\gamma=(t/T) \Delta t$. 

\begin{figure}[t]
    \centering
    \includegraphics[width=0.98\columnwidth]{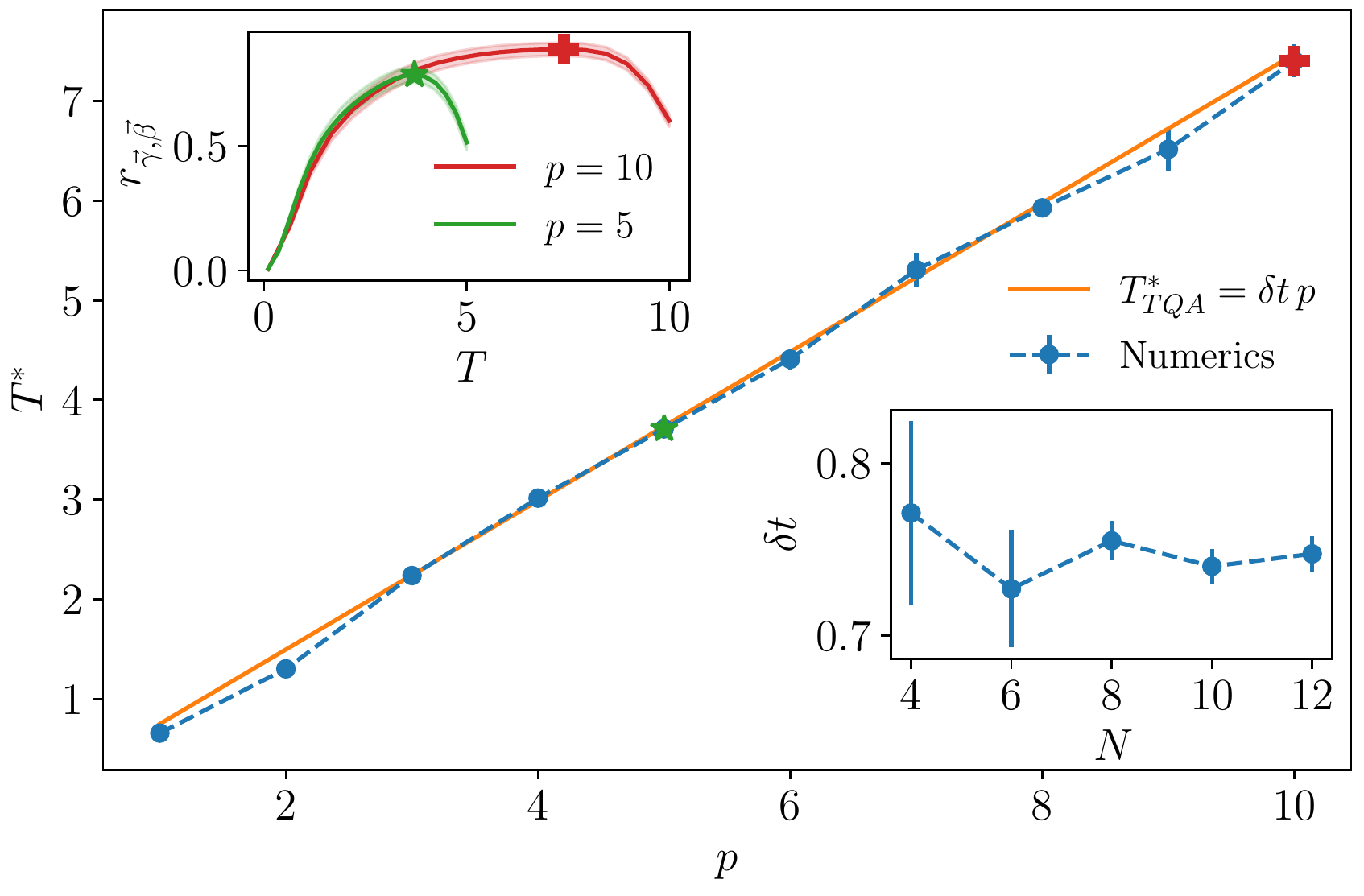}
    \caption{Optimal time of TQA evolution $T^*$ increases linearly with number of discretization steps $p$. Top inset illustrates that optimal performance of  TQA at time $T^*$ is followed by the rapid decrease in approximation ratio at longer times $T^*$. Data is shown for $N={12}$. Bottom inset shows finite size scaling of the time step $\slope$, determined by the slope of the $T^*$ vs $p$ dependence, that assumes approximately constant value with the graph size. All averaging is performed over 50 random instances of unweighted 3-regular graphs.}
    \label{fig3}
\end{figure}

Applying such decomposition to the quantum annealing protocol that is uniformly discretized on a grid of evolution times $t_i = i \Delta t$ with $i=1, ..., p$ and time step  $\Delta t = T/p$,  we obtain the unitary circuit equivalent to the depth-$p$ QAOA ansatz~(\ref{qaoa}) with angles being
\begin{equation}\label{Eq:TQA}
\gamma_i=\frac{i}{p} \Delta t,
\quad 
\beta_i=\left(1-\frac{i}{p}\right) \Delta t.
\end{equation}
In what follows we refer to such choice of angles as TQA initialization,  controlled by the time step $\Delta t$ at a fixed depth $p$. 

The mapping between TQA and QAOA along with the universality of quantum annealing for $T\to \infty$ was previously used as an argument for the existence of good QAOA protocols at depths $p\to \infty$~\cite{farhi2014quantum}. Typically the required evolution time of quantum annealing is inversely proportional to the square of the minimal energy gap $T\propto\Delta^{-2}$ encountered in the Hamiltonian $H(t)$ over the time evolution. Numerous studies established that the time required for a good performance often blows up exponentially due to the encounter of exponentially small gaps in $N$~\cite{albash2018adiabatic}.

In contrast to previous studies, we investigate TQA performance in a different setting that is motivated by its subsequent usage as a QAOA initialization.  The QAOA is characterized by a fixed circuit depth, $p$. Therefore, we fix $p$ and study the performance of TQA as a function of total time $T$ or, equivalently, time step $\Delta t$, related as $T = p\, \Delta t$. Generally the performance of quantum annealing tends to increase with the total annealing time. However in case of fixed $p$, longer annealing time corresponds to a coarser discretization, which leads to larger Trotter errors that scale proportionally to ${\cal O}(\Delta t^2)$ at small values of $\Delta t$. It is the interplay between increased efficiency and Trotter errors that leads to the existence of an optimal annealing time in the present setting. This is illustrated in Fig.~\ref{fig3} (top inset), where the approximation ratio for the TQA protocol increases with $T$ for small times, reaching a maximum at time $T^*$  followed by a sharp downturn. The sharp decrease of QA performance after $T^*$ was reported by \textcite{heyl2019quantum}, who attributed it to a phase transition caused by a proliferation of Trotter errors.

Main panel of Fig.~\ref{fig3} reveals a linear scaling of the optimal time $T^*$ with the number of time steps $p$. This is equivalent to the existence of an \emph{optimal time step $\slope$}, that determines $T^*$  as
\begin{equation}
    T^*_\text{TQA} = \slope \,p.
\end{equation}
The bottom inset in Fig.~\ref{fig3} shows that the time step $\slope$ defined as a slope of a linear fit of $T^*$ with $p$ converges with the problem size $N$. This gives a strong evidence that $\slope$ is a well-defined quantity in the thermodynamic limit $N\to\infty$. For the family of the 3-regular graphs considered here we observe that the optimal time step tends to value $\slope\approx 0.75$. The existence of an optimal time step that is of order one holds for three other graph ensembles, considered in Appendix~\ref{app2}, although the numerical value of this time step depends on the specific graph ensemble.  

We use the TQA initialization in Eq.~(\ref{Eq:TQA}) with time step $\Delta t=0.75$ for the QAOA and observe in Fig.~\ref{fig2} that it allows to avoid the local minima and helps the QAOA to converge to a minimum that is very close to the global minimum in terms of approximation ratio. This result motivates the systematic analysis of the  performance of the TQA initialization.

\subsection{TQA initialization of QAOA}

\begin{figure}[t]
    \centering
    \includegraphics[width=\columnwidth]{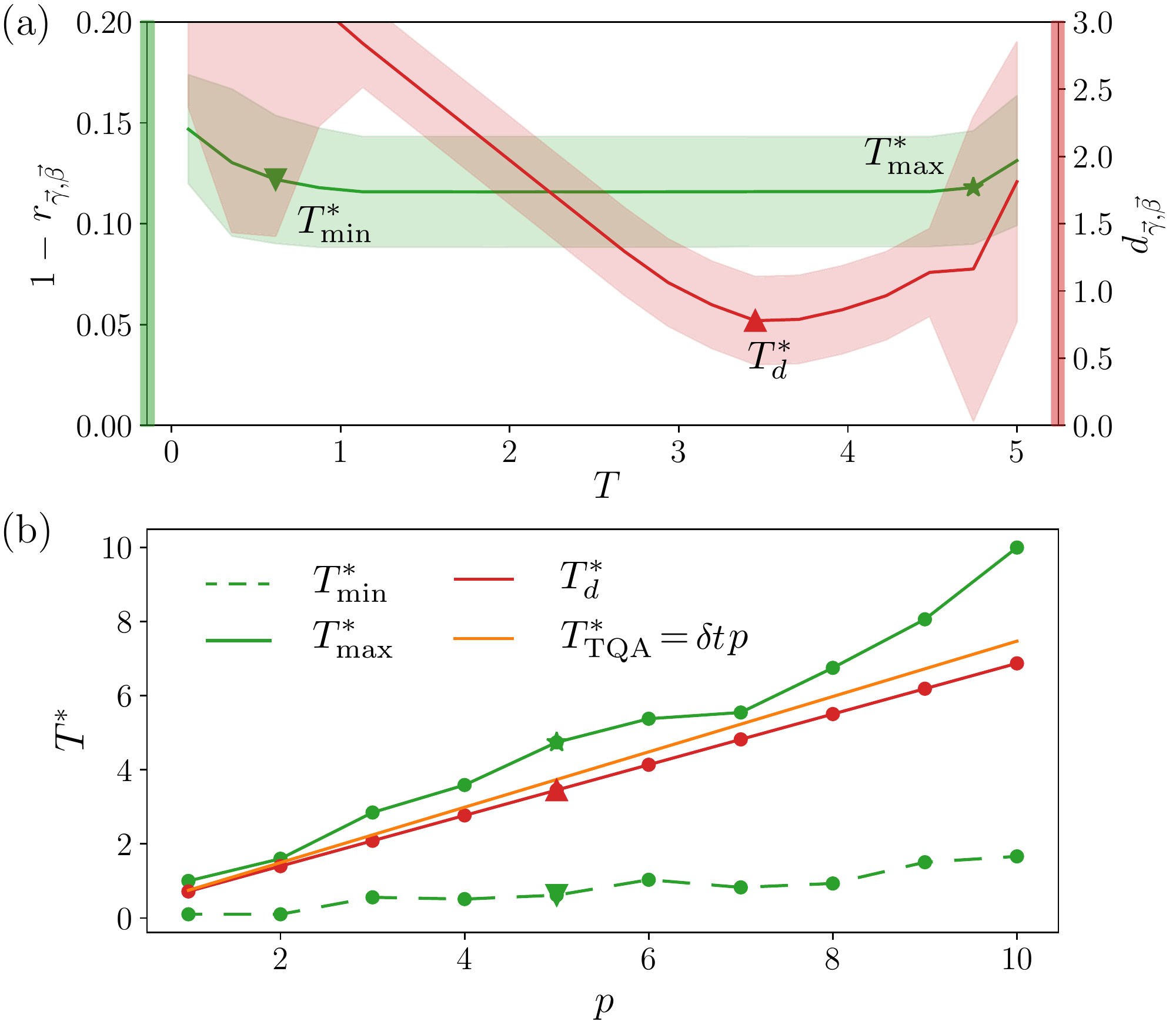}
    \caption{
(a) Approximation ratio of the $p=5$ QAOA as a function of TQA initialization time $T$ reveals that a range of initialization times $[T^*_\text{min},T^*_\text{max}]$ (green triangle and star) yield the performance within $1\%$ of the minimal $1-r_{\vec{\gamma}, \vec{\beta}}$. On the other hand, the study of the distance between the TQA initialization and the converged value of the angles reveals the existence of a time $T^*_d$ where the QAOA performs the smallest parameter updates. 
(b) All three times $T^*_\text{min}$, $T^*_\text{max}$, and $T^*_d$ defined in panel (a) increase linearly with QAOA circuit depth $p$. Moreover, $T^*_d$ is very close to the time where the TQA protocol itself achieves optimal performance, $T^*_\text{TQA}$, see Fig.~\ref{fig2}. Data was obtained for $N=12$ and averaged over 50 random graphs.
}
    \label{fig4}
\end{figure}

We continue with a detailed study of the TQA initialization defined in Eq.~(\ref{Eq:TQA}) as a function of time $T$ at fixed $p$. The green line in Fig.~\ref{fig4}(a) reveals that the approximation ratio remains constant for a range of times, denoted as $[T^*_\text{min},T^*_\text{max}]$. This figure shows results for $p=5$ QAOA applied to graphs with $N=12$ vertices, but a similar trend holds for other values of depth, problem sizes, and graph ensembles. The constant approximation ratio in a range of $T$ is naturally explained by the convergence of parameter optimization routine to the same minimum for $T\in [T^*_\text{min},T^*_\text{max}]$, see cartoon in Fig.~\ref{fig1}(c). In order to discriminate between different times in the above range, we study the distance between initialization parameters and optimized values of $\vec\gamma,\vec\beta$. The red line Fig.~\ref{fig4}(a) shows that this  distance has a well-pronounced minimum at a time denoted as $T^*_d$ that is contained within the same interval $[T^*_\text{min},T^*_\text{max}]$. The TQA initialization with time $T^*_d$ is closest to the local minimum achieved from it in a sense of distance defined in Eq.~(\ref{angular_distance}).

All three times $T^*_\text{min}$, $T^*_\text{max}$, and $T^*_d$ were defined above using the QAOA with fixed depth $p$. Figure~\ref{fig4}(b) reveals that all three times scale approximately linearly with $p$. This allows to define a range of time steps for the TQA initialization that yield the same performance of optimized QAOA, $\Delta t \in [{ 0.16},{ 0.92}]$ for the present graph ensemble. Moreover, the time  $T^*_d$ nearly coincides with the optimal TQA time $T^*_\text{TQA} = \slope\,p$ obtained in the previous section, implying that $\Delta t =\delta t = 0.75$ is the optimal value of time step. This result also holds for the MaxCut problem on other graph families, see Appendix. 

The similarity between the optimal time of the TQA protocol to the time where the angular distance $d_{\vec{\gamma}, \vec{\beta}}$ between the initial and final protocol is minimized, suggests that the performance of the QAOA is bounded by the same phase transition that occurs in TQA~\cite{heyl2019quantum}. However, the QAOA is able to provide a significant improvement over TQA by doing additional optimizations of variational parameters. Recent work~\cite{zhou2018quantum} suggested that such performance improvement may be due to utilization of ``diabatic pumps'' that allow to return the population from excited states back to the ground state. This could potentially explain the systematic deviation of the QA protocol from TQA initialization as seen in Fig.~\ref{app_fig3} in Appendix~\ref{app3}.

\begin{figure}[t]
    \centering
    \includegraphics[width=\columnwidth]{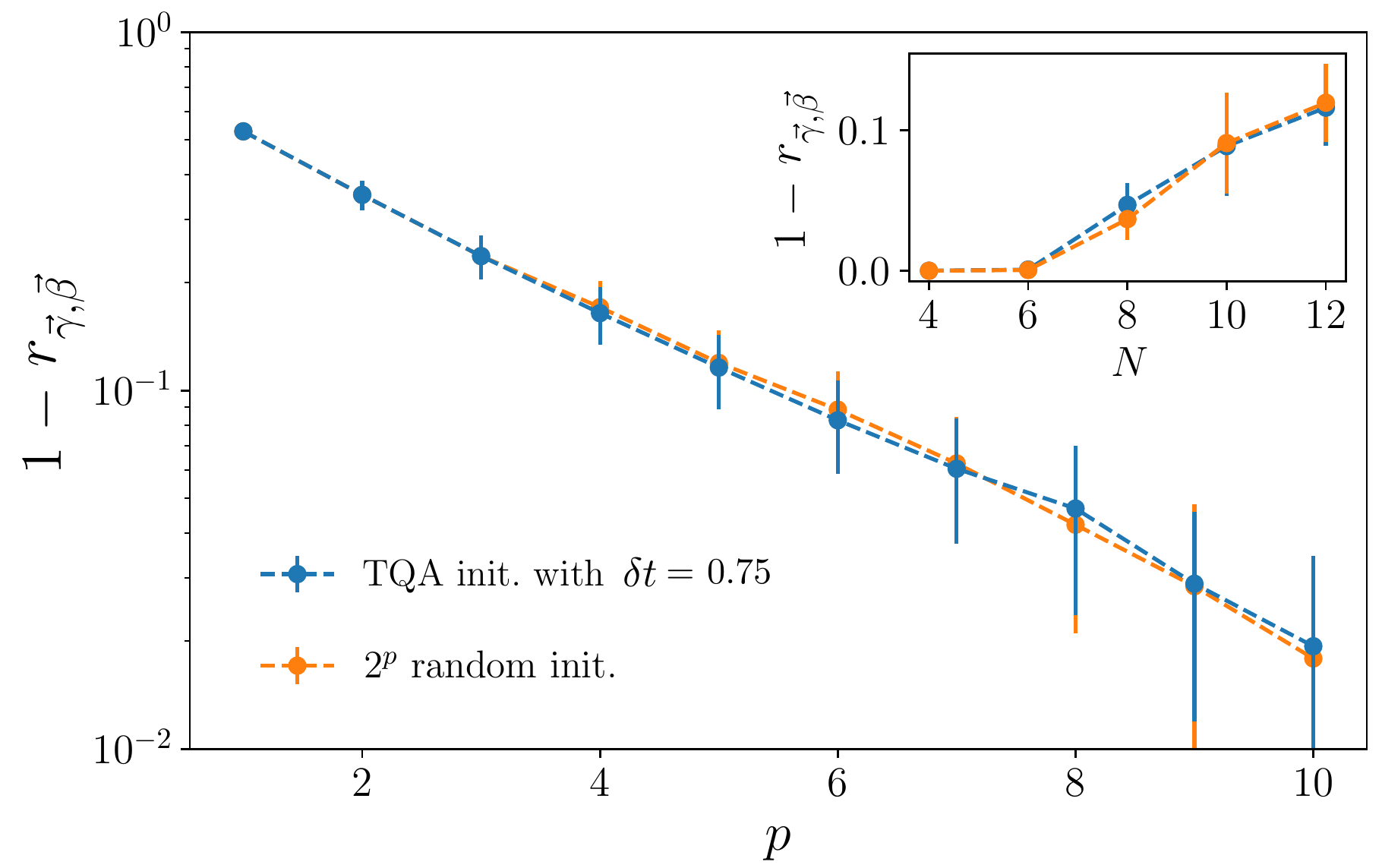}
    \caption{ A single optimization run of the QAOA with TQA initialization with time $T=\slope\,p$  yields equivalent performance to the best out of $2^p$ random initializations. System size is $N=12$. Inset reveals that the comparable performance persists over the entire range of considered system sizes, circuit depth is $p=10$. Averaging was performed over 50 random graphs.}
    \label{fig5}
\end{figure}

Finally, we compare the performance of QAOA that used $2^p$ random initializations to the QAOA launched from TQA initialization with optimal time step $\slope$. Surprisingly, Fig.~\ref{fig5} shows that TQA initialization yields the \emph{same} performance as the best result for random initialization even for QAOA protocols with depth comparable to the problem size, $N$. Moreover, the inset of Fig.~\ref{fig5} illustrates that the excellent performance of TQA initialization holds true for a broad range of system sizes $N$, while Appendix~\ref{app4} presents equally encouraging results for other graph ensembles. Note that the QAOA performance for fixed $p$ decreases with system size $N$, which was attributed to the fact that the QAOA with fixed $p$ cannot ``probe'' the whole graph. In order for the QAOA to achieve constant performance for increasing problem size $N$, the depth of QAOA should increase at least as~$\log N$~\cite{zhou2018quantum}. 

\section{Summary and discussion}\label{discussion}

Our central result is the establishment of a family of TQA initializations for the QAOA parametrized by a time step $\Delta t$. We find that TQA initialization allows the QAOA to find a solution close to the global optima for a broad range of parameter $\Delta t$. In this range our initialization scheme achieves results similar to the best outcome of $2^p$ random initializations, with a single optimization run. Moreover we establish a heuristic way to identify the optimal $\Delta t$ for the TQA initialization from the performance of the TQA protocol. 

Our results open the door to more time-efficient practical implementations of the QAOA on NISQ devices. To this end, we propose a two-step practical NISQ algorithm that capitalizes on the success of TQA initialization and uses the heuristic results to establish an optimal value of the time step. The first two steps of Algorithm~\ref{algorithm} implement  the TQA protocol on a NISQ device, thus obtaining an estimate for the optimal time in the TQA initialization. This can be readily carried out on today's NISQ devices~\cite{smith2019simulating}. The second part of the algorithm consists of running the QAOA optimization loop using values of variational parameters according to Eq.~(\ref{Eq:TQA}).

\begin{algorithm}[H]
    \caption{QAOA with TQA initialization}\label{algorithm}
    \begin{algorithmic}[1]
        \State Implement QAOA ansatz with circuit depth $p$.
        \State Estimate  time step $\slope$ using TQA:\newline 
       find optimal time $T^* \leftarrow \arg \min_{T} \big<H_C\big>_{p}$\newline
       and set $\slope\leftarrow\frac{T^*}{p}$, see Fig.~\ref{fig3}.
        \State Use TQA initialization $\gamma_i \leftarrow \frac{i}{p} \slope$ and $\beta_i\leftarrow(1-\frac{i}{p}) \slope$.
        \State Run the QAOA parameter optimization, see Fig.~\ref{fig1}.
    \end{algorithmic}
\end{algorithm}

Numerical simulations presented above suggest good performance of the above algorithm in the idealized case when presence of noise, gate errors, and other imperfections are neglected. Moreover, the fact that TQA initialization converges to a good minimum for the range of times (equivalently, time steps) $T\in [T^{*}_{\min}, T^{*}_{\max}]$, see Fig.~\ref{fig4}, suggests that this algorithm has a high tolerance towards imperfections in determining the value of $\delta t$. Determining the performance of this algorithm on a real NISQ device or incorporating some of the imperfections into the numerical simulation remains an interesting open problem. 

In our studies we restricted our attention to the MaxCut problem and demonstrated success of our approach for three different random graph ensembles. We expect that these results also hold for other graph ensembles, provided that the concentration of the QAOA landscape is true~\cite{Brandao}. It is also interesting to check if our findings hold true beyond the MaxCut problem. Furthermore, it will be interesting to study the finite size scaling for problem sizes $N>12$ considered here using matrix product states~\cite{schollwock2011density} or neural-network quantum states~\cite{carleo2017solving,medvidovic2020classical}. 

In addition to practical NISQ algorithms, our finding suggest a previously unknown connection between the QAOA at relatively small circuit depth and quantum annealing. The fact that quantum annealing inspired initializations belong to a basin of attraction of a high-quality minimum in the QAOA landscape, see Fig.~\ref{fig1}(c), invites a more comprehensive study of the QAOA landscape from this perspective. How many good quality minima typically exist in such landscape? How different are they from each other and what are their basins of attraction? Can one use other information measures such as entanglement or Fisher information~\cite{abbas2020power} to characterize the QAOA landscape? Finding answers to such questions may lead to other prospective families of QAOA initializations.

While TQA provides a good initialization, the subsequent QAOA optimization is able to significantly improve the performance. Understanding the underlying mechanisms of such performance improvement is an outstanding challenge. In particular, there remains an intriguing possibility that the QAOA optimization routine implements some of the techniques, developed to improve the annealing fidelity, such as diabatic pumps~\cite{zhou2018quantum}, shortcuts to adiabaticity~\cite{STA}, and counterdiabatic driving~\cite{Sels,Clayes}. The fact that the optimal time step coincides with the point of proliferation of Trotter errors~\cite{heyl2019quantum}, thus effectively taking maximal possible value suggests interesting parallels to the Pontryagin's minimum principle considered in context of variational quantum algorithms~\cite{Chamon}.

To conclude, we hope that TQA initialization of the QAOA established in this work will help to achieve practical quantum advantage by executing the QAOA on available devices and inspire future research that could lead to better understanding of what happens under the hood of QAOA optimization.

\section*{Data and code availability}
Data is available upon reasonable request, a brief tutorial for the TQA initialization can be found in Ref.~\cite{stefan}

\section*{Acknowledgments}
We would like to thank D. Abanin and  R. Medina for fruitful discussions and A. Smith and I. Kim for valuable feedback on the manuscript. We acknowledge support by the European Research Council (ERC) under the European Union's Horizon 2020 research and innovation program (Grant Agreement No.~850899).
\appendix
\section{Optimization landscape for different graph ensembles}\label{app1}
We start by reviewing all graph ensembles used in the main text and Appendices. In particular, we focus on  symmetries that allow to reduce the space of QAOA parameters. 

{\it 3-regular unweighted graphs} represent the graph ensemble considered in the main text. Each vertex is connected exactly to three other vertices chosen at random. In order to sample graphs from this ensemble we use the  \texttt{networkx} Python package~\cite{networkx}. For 3-regular unweighted graphs the space of variational parameters can be restricted using the fact that the classical Hamiltonian has integer eigenvalues (thus $\gamma_i$ are defined modulo $\pi$) and that shifting any of angles $\beta_i$ by $\pi/2$ is equivalent to a spin flip of $H_C$ that has no effect~\cite{zhou2018quantum}. This allows to restrict  $\beta_i \in [-\frac{\pi}{4}, \frac{\pi}{4})$ and $\gamma_i \in [-\frac{\pi}{2}, \frac{\pi}{2})$, and is reflected in the definition of distance in Eq.~(\ref{angular_distance}) in the main text. 

{\it 3-regular weighted graphs} are characterized by presence of random weights $w_{ij}$ assigned to each edge $\langle i, j\rangle$. These weights are chosen to be $w_{ij}\in[0,1)$. Presence of random weights does not allow to restrict the domain of $\gamma_i$ angles as before, though restriction $\beta_i \in [-\frac{\pi}{4}, \frac{\pi}{4})$ still works. Therefore the analogue of Eq.~(\ref{angular_distance}) for this and other weighted ensembles reads $ d^{(w)}_{\vec \gamma,\vec \beta} = \sum_{i=1}^p  (|\beta_i - \beta^*_i|_{\frac{\pi}{2}} + |\gamma_i - {\gamma^*_i}|)$.

{\it Erd\H{o}s-R\'enyi  graphs} represent a random graph ensemble where two edges are connected on random with a fixed probability, chosen to be $q = 0.5$. In contrast to above examples, the  fixed value of $q$ implies that edge connectivity increases with number of vertices as $qN$. Erd\H{o}s-R\'enyi  graphs exhibit the same symmetries as 3-regular unweighted graphs.

\begin{figure}[t]
    \centering
    \includegraphics[width=\columnwidth]{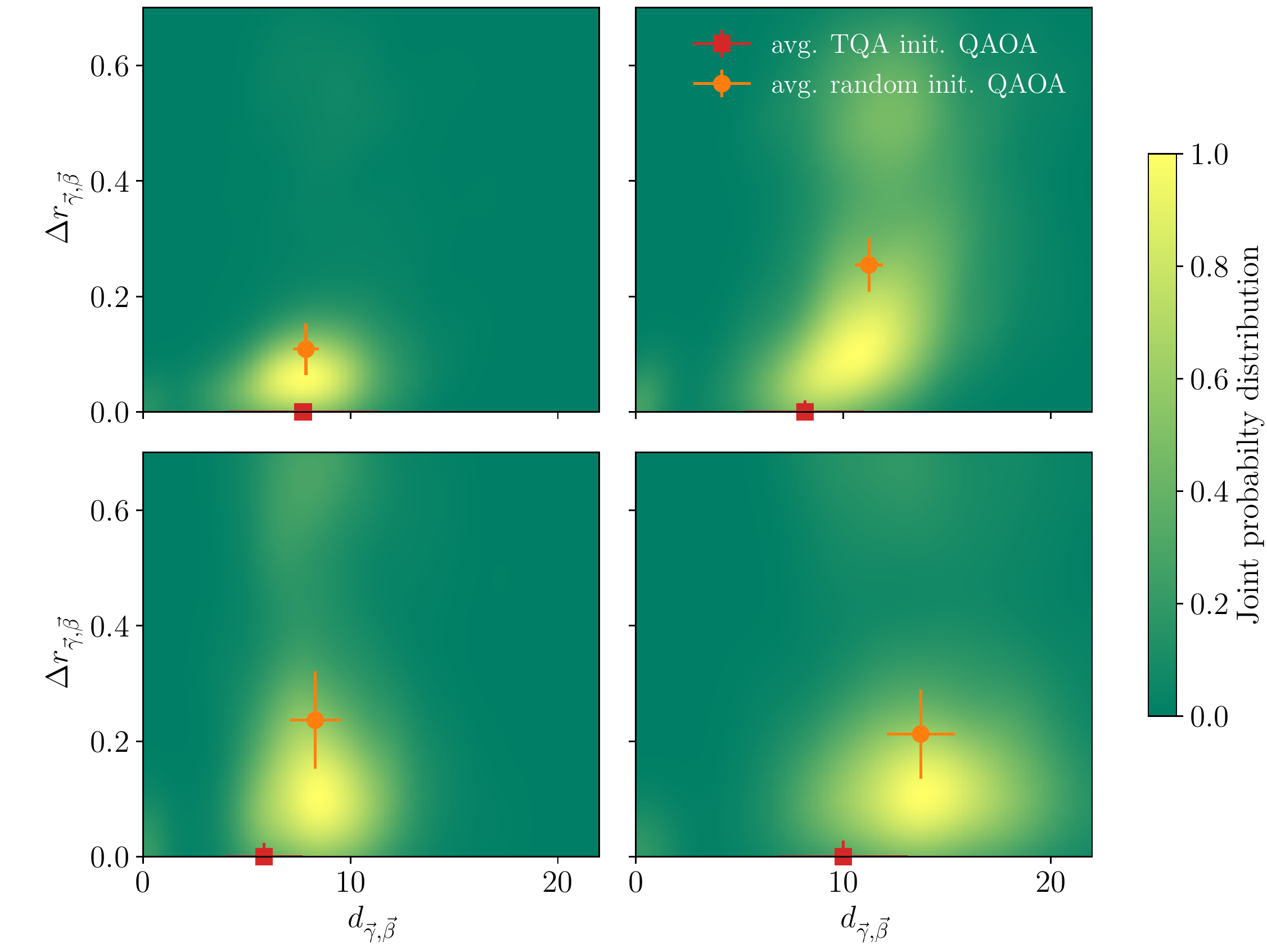}
    \caption{Comparing the joint probability distribution of the distance to the global minimum in parameter space $d_{\vec\gamma,\vec\beta}$ and in terms of approximation ratio $\Delta r_{\vec\gamma,\vec\beta}$ for weighted 3-regular (top) and Erd\H{o}s-R\'enyi graphs with edge probability $0.5$ (bottom) reveals that the distribution is dependent on the initialization interval  for weighted 3-regular graphs. We initialize the parameters for $k=1$ (left) and $k=2$ (right) and observe that for weighted 3-regular graphs the enlarged interval leads to an increased spread of the local optimas in $\Delta r_{\vec\gamma,\vec\beta}$ (yellow region). The spread in $\Delta r_{\vec\gamma,\vec\beta}$ for Erd\H{o}s-R\'enyi graphs remains largely unaffected, as expected from the symmetry considerations. Similarly to Fig.~\ref{fig2}, red squares correspond to the QAOA minimum achieved from TQA initialization (shifted from small negative values of $\Delta r_{\vec\gamma,\vec\beta}$ to zero for improved visibility), orange dots correspond to the average performance of random initialization. Data is for 50 random graphs with $N=10$ and $p=5$.}
    \label{app_fig1}
\end{figure}

The presence of an unbounded region of parameters $\gamma_i$ in the weighted graph ensemble represents an additional challenge in visualizing the QAOA optimization landscape and choice of initialization parameter. In order to explore the importance of large values of $|\gamma_i|$, we consider the sequence of enlarged intervals $\gamma_i \in [-k\frac{\pi}{2}, k\frac{\pi}{2})$ with $k=1,2$.  Figure~\ref{app_fig1} shows the joint probability distributions similar to Fig.~\ref{fig2}. We see that for 3-regular weighted graphs the enlarged initialization interval $k=2$ leads to a concentration of local optima further away from the global solution compared to the $k=1$ interval. When we repeat the same analysis for Erd\H{o}s-R\'enyi graphs, we observe that $\Delta r_{\vec{\gamma}, \vec{\beta}}$ is unaffected by the enlarged $k=2$ interval. This numerically confirms the symmetry considerations from above and allows us to restrict $\vec{\gamma}$ to the $k=1$ interval in all further analysis. For unweighted graphs such restriction relies on symmetry,  and for weighted graphs this is motivated by the fact that an extended region of $\gamma_i$ worsens the performance of random initialization in the QAOA.

\section{Optimal time for TQA}\label{app2}

Below we  discuss the dependence of the optimal time step $\slope$ of the TQA algorithm on the graph ensemble. An analytical \emph{upper bound} on the number of Trotter steps $p$ needed to approximate the time evolution with precision $\epsilon$ in terms of operator trace distance was obtained in Ref.~\cite{berry2007efficient}. Translating this bound into the scaling of $\slope$ we obtain  $\slope \propto 1/({||H_C||_F N})$, where $||H_C||_F$ is the Frobenius norm of the classical Hamiltonian. This norm exponentially diverges with $N$, suggesting very small values of $\slope$ at large system sizes. This is not surprising, since the bound of Ref.~\cite{berry2007efficient} operates on the distance between two many-body unitary operators. In contrast, the performance of the TQA algorithm is studied using the approximation ratio that quantifies how close the expectation value of the local observable $H_C$, is to the ground state energy.

\begin{figure}[t]
    \centering
    \includegraphics[width=\columnwidth]{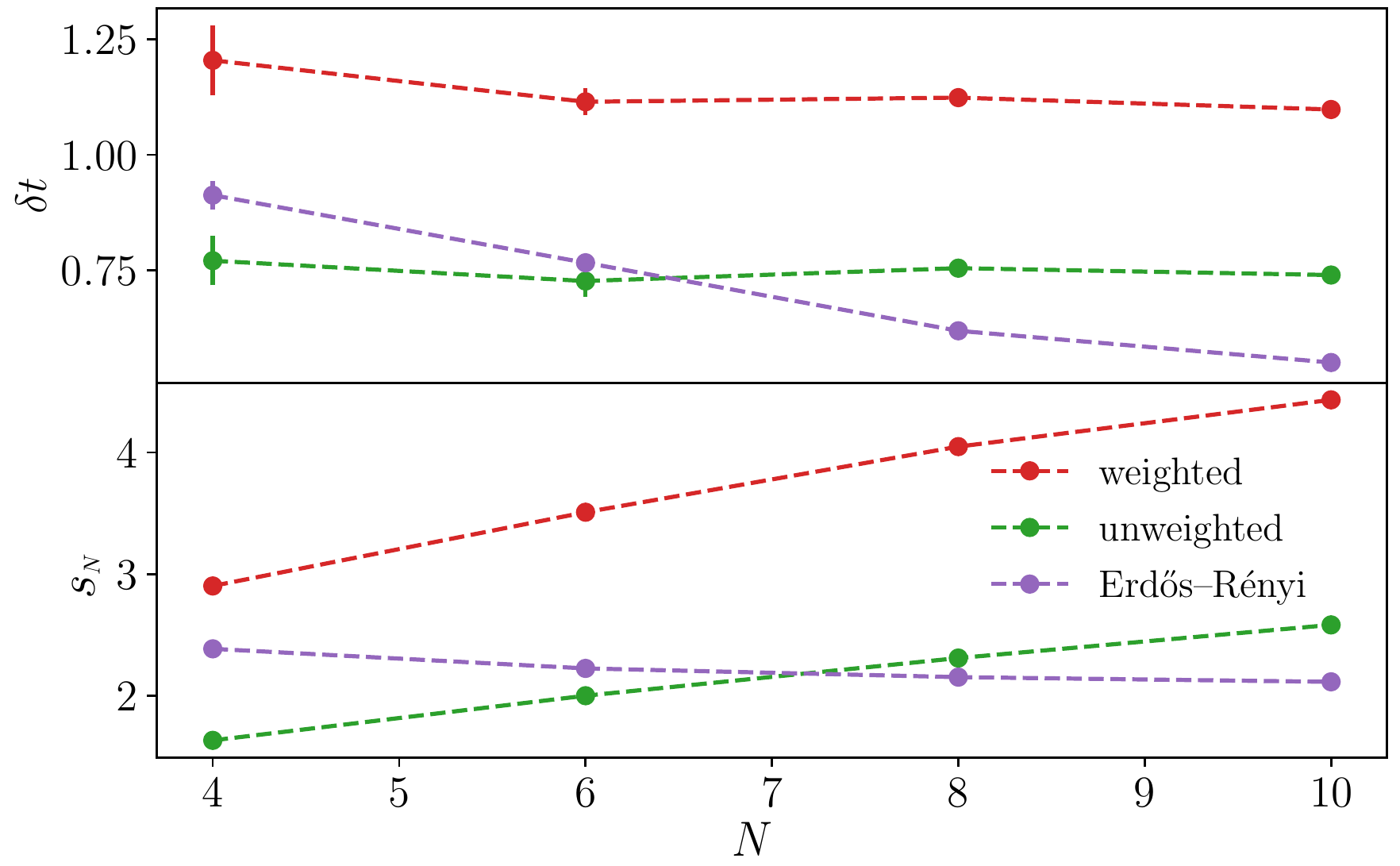}
    \caption{(Top) Optimal time time step of TQA evolution $\slope$ is largely independent of system size and scales qualitatively similar to Eq.~(\ref{kappa_scaling}) shown in the bottom panel.}
    \label{app_fig2}
\end{figure}

The effect of Trotterization on local observables was considered in Ref.~\cite{heyl2019quantum}. This work conjectured the existence of  a finite value of the time step of order one, at which the discretization of time evolution fails to approximate the local observables. This value of the time step may be related to the convergence radius of the Baker-Campbell-Hausdorff series expansion, which is governed by the norm of the classical Hamiltonian and its commutator with $H_B$. Phenomenologically, the Frobenius norm divided by the square root of Hilbert space dimension and problem size $N$, 
\begin{equation} \label{kappa_scaling}
    s_N = \frac{N 2^{N/2}}{||H_C||_{F}},
\end{equation}
is expected to be $N$-independent in  the thermodynamic limit. 

\begin{figure}[t]
    \centering
    \includegraphics[width=0.95\columnwidth]{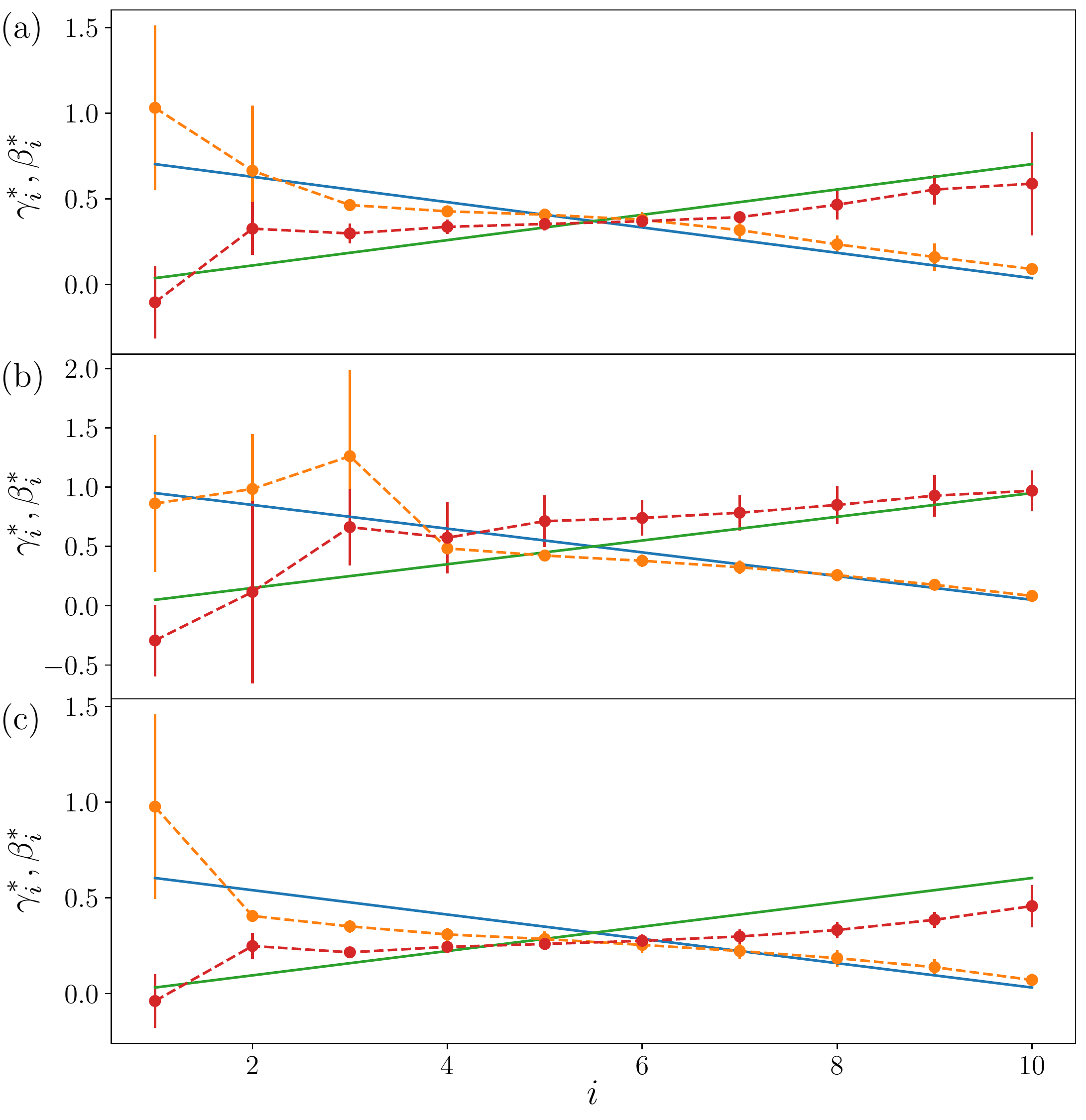}
    \caption{Converged parameters $\vec{\gamma}^*$ (red) and $\vec{\beta}^*$ (orange) show only slight alterations from the TQA initialization indicated by the green and blue lines respectively. The QAOA optimization modifies parameters at small $i$, while they remain TQA-like in the rest of the protocol. The results were averaged over 50 random unweighted 3-regular graphs (a), weighted 3-regular graphs (b) and  Erd\H{o}s-R\'enyi  graphs (c), all data is for $p=10$ and $N=10$.}
    \label{app_fig3}
\end{figure}

Figure~\ref{app_fig2} compares the dependence of $\slope$ on the system size with the phenomenological scaling  $s_N$ defined in Eq.~(\ref{kappa_scaling}). We observe that the expression $s_N$ qualitatively matches the numerical scaling that we observe for $\slope$ between different graph ensembles. In particular, the value of the time step is largest for weighted 3-regular graphs that are expected to have the smallest norm of the classical Hamiltonian.  However, $s_N$ fails to capture $\slope$ quantitatively, highlighting the need to develop a better analytical understanding of the point that governs the phase transition from localization to quantum chaos for local observables according to Ref.~\cite{heyl2019quantum}.

\begin{figure}[t]
    \centering
    \includegraphics[width=0.95\columnwidth]{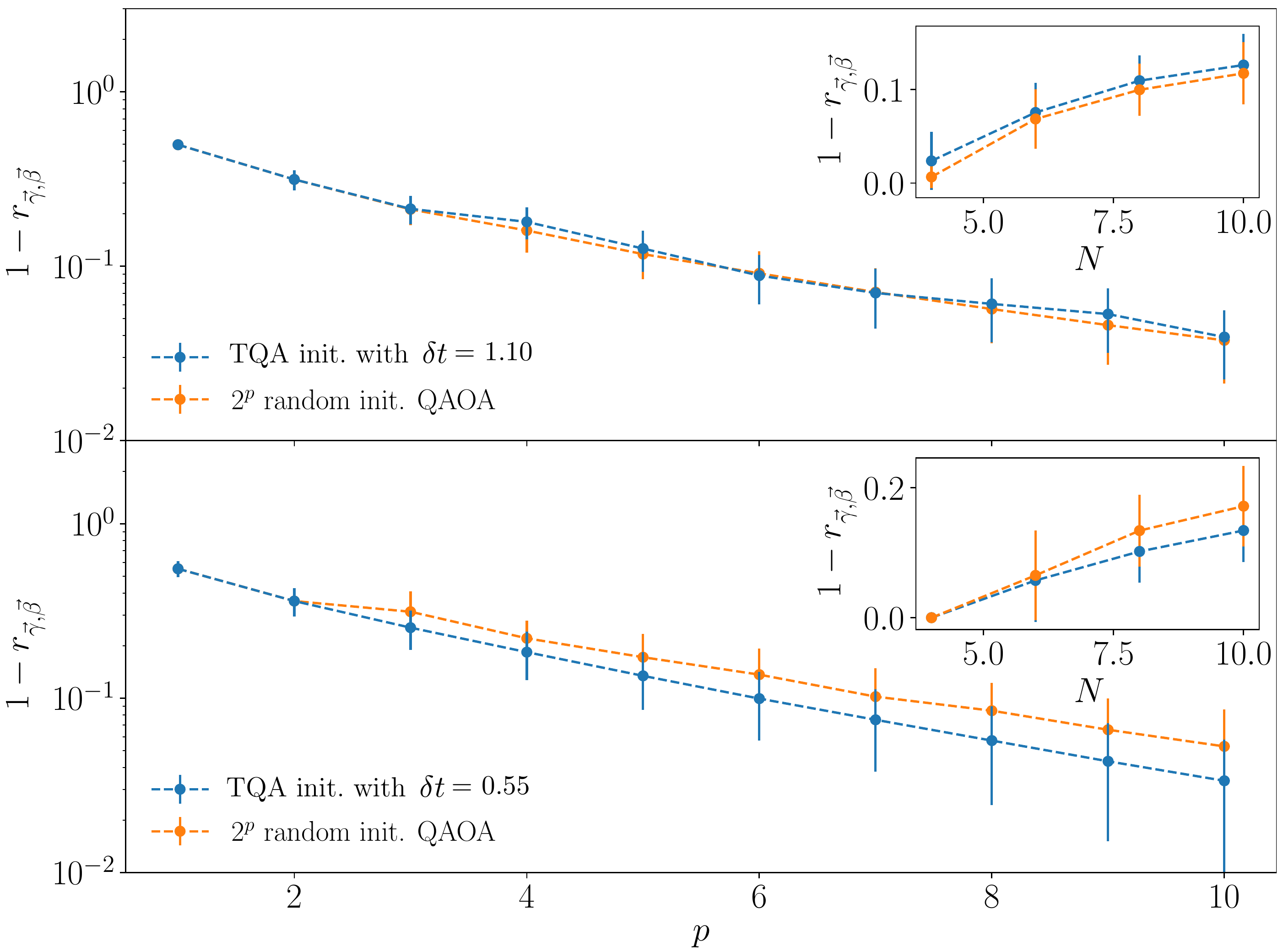}
    \caption{TQA initialization leads to the same QAOA performance as the best of $2^p$ random initializations for both weighted 3-regular graphs (top) and Erd\H{o}s-R\'enyi graphs (bottom). We average the results over 50 graph realizations, the main plot was obtained for system size $N=10$, inset is for circuit depth $p=10$.}
    \label{app_fig4}
\end{figure}

\section{Patterns in optimized parameters}\label{app3}
The QAOA is inspired by TQA and is thus universal for $p\rightarrow \infty$. However, for finite $p$ the converged QAOA parameters also display stark similarity to a QA protocol which was noticed in some earlier works~\cite{zhou2018quantum, crooks2018performance}. In Fig.~\ref{app_fig3} we compare the TQA initialization and final QAOA parameters. The QAOA parameters show only slight alterations at the beginning of the protocol and remain close to their original values throughout the rest of the protocol. This holds true for the three graph types that we considered in our analysis. In addition, the small variation between optimal parameters for different graph instances is in line with the concentration of the QAOA landscape demonstrated analytically at low $p$ in Ref.~\cite{Brandao}.

\section{Random vs TQA initialization for other graph ensembles}\label{app4}
In addition to the unweighted 3-regular graphs, discussed in the main text, we also test TQA initialization on weighted 3-regular graphs and Erd\H{o}s-R\'enyi graphs. We find that TQA initialization yields the same performance as the best of random initializations for weighted 3-regular graphs, see Fig.~\ref{app_fig4}. For Erd\H{o}s-R\'enyi, TQA initialization even outperforms the best of $2^p$ random initializations.


\begin{thebibliography}{48}%
\makeatletter
\providecommand \@ifxundefined [1]{%
 \@ifx{#1\undefined}
}%
\providecommand \@ifnum [1]{%
 \ifnum #1\expandafter \@firstoftwo
 \else \expandafter \@secondoftwo
 \fi
}%
\providecommand \@ifx [1]{%
 \ifx #1\expandafter \@firstoftwo
 \else \expandafter \@secondoftwo
 \fi
}%
\providecommand \natexlab [1]{#1}%
\providecommand \enquote  [1]{``#1''}%
\providecommand \bibnamefont  [1]{#1}%
\providecommand \bibfnamefont [1]{#1}%
\providecommand \citenamefont [1]{#1}%
\providecommand \href@noop [0]{\@secondoftwo}%
\providecommand \href [0]{\begingroup \@sanitize@url \@href}%
\providecommand \@href[1]{\@@startlink{#1}\@@href}%
\providecommand \@@href[1]{\endgroup#1\@@endlink}%
\providecommand \@sanitize@url [0]{\catcode `\\12\catcode `\$12\catcode
  `\&12\catcode `\#12\catcode `\^12\catcode `\_12\catcode `\%12\relax}%
\providecommand \@@startlink[1]{}%
\providecommand \@@endlink[0]{}%
\providecommand \url  [0]{\begingroup\@sanitize@url \@url }%
\providecommand \@url [1]{\endgroup\@href {#1}{\urlprefix }}%
\providecommand \urlprefix  [0]{URL }%
\providecommand \Eprint [0]{\href }%
\providecommand \doibase [0]{https://doi.org/}%
\providecommand \selectlanguage [0]{\@gobble}%
\providecommand \bibinfo  [0]{\@secondoftwo}%
\providecommand \bibfield  [0]{\@secondoftwo}%
\providecommand \translation [1]{[#1]}%
\providecommand \BibitemOpen [0]{}%
\providecommand \bibitemStop [0]{}%
\providecommand \bibitemNoStop [0]{.\EOS\space}%
\providecommand \EOS [0]{\spacefactor3000\relax}%
\providecommand \BibitemShut  [1]{\csname bibitem#1\endcsname}%
\let\auto@bib@innerbib\@empty
\bibitem [{\citenamefont {{Arute}}\ \emph
  {et~al.}(2020{\natexlab{a}})\citenamefont {{Arute}} \emph
  {et~al.}}]{arute2020quantum}%
  \BibitemOpen
  \bibfield  {author} {\bibinfo {author} {\bibfnamefont {F.}~\bibnamefont
  {{Arute}}} \emph {et~al.},\ }\bibfield  {title} {\bibinfo {title} {{Quantum
  Approximate Optimization of Non-Planar Graph Problems on a Planar
  Superconducting Processor}},\ }\href@noop {} {\bibfield  {journal} {\bibinfo
  {journal} {arXiv e-prints}\ ,\ \bibinfo {eid} {arXiv:2004.04197}} (\bibinfo
  {year} {2020}{\natexlab{a}})},\ \Eprint {https://arxiv.org/abs/2004.04197}
  {arXiv:2004.04197 [quant-ph]} \BibitemShut {NoStop}%
\bibitem [{\citenamefont {{Arute}}\ \emph
  {et~al.}(2020{\natexlab{b}})\citenamefont {{Arute}} \emph
  {et~al.}}]{arute2020hartree}%
  \BibitemOpen
  \bibfield  {author} {\bibinfo {author} {\bibfnamefont {F.}~\bibnamefont
  {{Arute}}} \emph {et~al.},\ }\bibfield  {title} {\bibinfo {title}
  {{Hartree-Fock on a superconducting qubit quantum computer}},\ }\href
  {https://doi.org/10.1126/science.abb9811} {\bibfield  {journal} {\bibinfo
  {journal} {Science}\ }\textbf {\bibinfo {volume} {369}},\ \bibinfo {pages}
  {1084} (\bibinfo {year} {2020}{\natexlab{b}})},\ \Eprint
  {https://arxiv.org/abs/2004.04174} {arXiv:2004.04174 [quant-ph]} \BibitemShut
  {NoStop}%
\bibitem [{\citenamefont {{Arute}}\ \emph
  {et~al.}(2020{\natexlab{c}})\citenamefont {{Arute}} \emph
  {et~al.}}]{arute2020observation}%
  \BibitemOpen
  \bibfield  {author} {\bibinfo {author} {\bibfnamefont {F.}~\bibnamefont
  {{Arute}}} \emph {et~al.},\ }\bibfield  {title} {\bibinfo {title}
  {{Observation of separated dynamics of charge and spin in the Fermi-Hubbard
  model}},\ }\href@noop {} {\bibfield  {journal} {\bibinfo  {journal} {arXiv
  e-prints}\ ,\ \bibinfo {eid} {arXiv:2010.07965}} (\bibinfo {year}
  {2020}{\natexlab{c}})},\ \Eprint {https://arxiv.org/abs/2010.07965}
  {arXiv:2010.07965 [quant-ph]} \BibitemShut {NoStop}%
\bibitem [{\citenamefont {{Wright}}\ \emph {et~al.}(2019)\citenamefont
  {{Wright}} \emph {et~al.}}]{wright2019benchmarking}%
  \BibitemOpen
  \bibfield  {author} {\bibinfo {author} {\bibfnamefont {K.}~\bibnamefont
  {{Wright}}} \emph {et~al.},\ }\bibfield  {title} {\bibinfo {title}
  {{Benchmarking an 11-qubit quantum computer}},\ }\href
  {https://doi.org/10.1038/s41467-019-13534-2} {\bibfield  {journal} {\bibinfo
  {journal} {Nature Communications}\ }\textbf {\bibinfo {volume} {10}},\
  \bibinfo {eid} {5464} (\bibinfo {year} {2019})},\ \Eprint
  {https://arxiv.org/abs/1903.08181} {arXiv:1903.08181 [quant-ph]} \BibitemShut
  {NoStop}%
\bibitem [{\citenamefont {{Preskill}}(2018)}]{preskill2018quantum}%
  \BibitemOpen
  \bibfield  {author} {\bibinfo {author} {\bibfnamefont {J.}~\bibnamefont
  {{Preskill}}},\ }\bibfield  {title} {\bibinfo {title} {{Quantum Computing in
  the NISQ era and beyond}},\ }\href@noop {} {\bibfield  {journal} {\bibinfo
  {journal} {arXiv e-prints}\ ,\ \bibinfo {eid} {arXiv:1801.00862}} (\bibinfo
  {year} {2018})},\ \Eprint {https://arxiv.org/abs/1801.00862}
  {arXiv:1801.00862 [quant-ph]} \BibitemShut {NoStop}%
\bibitem [{\citenamefont {{Farhi}}\ \emph {et~al.}(2014)\citenamefont
  {{Farhi}}, \citenamefont {{Goldstone}},\ and\ \citenamefont
  {{Gutmann}}}]{farhi2014quantum}%
  \BibitemOpen
  \bibfield  {author} {\bibinfo {author} {\bibfnamefont {E.}~\bibnamefont
  {{Farhi}}}, \bibinfo {author} {\bibfnamefont {J.}~\bibnamefont
  {{Goldstone}}},\ and\ \bibinfo {author} {\bibfnamefont {S.}~\bibnamefont
  {{Gutmann}}},\ }\bibfield  {title} {\bibinfo {title} {{A Quantum Approximate
  Optimization Algorithm}},\ }\href@noop {} {\bibfield  {journal} {\bibinfo
  {journal} {arXiv e-prints}\ ,\ \bibinfo {eid} {arXiv:1411.4028}} (\bibinfo
  {year} {2014})},\ \Eprint {https://arxiv.org/abs/1411.4028} {arXiv:1411.4028
  [quant-ph]} \BibitemShut {NoStop}%
\bibitem [{\citenamefont {Zhou}\ \emph {et~al.}(2020)\citenamefont {Zhou},
  \citenamefont {Wang}, \citenamefont {Choi}, \citenamefont {Pichler},\ and\
  \citenamefont {Lukin}}]{zhou2018quantum}%
  \BibitemOpen
  \bibfield  {author} {\bibinfo {author} {\bibfnamefont {L.}~\bibnamefont
  {Zhou}}, \bibinfo {author} {\bibfnamefont {S.-T.}\ \bibnamefont {Wang}},
  \bibinfo {author} {\bibfnamefont {S.}~\bibnamefont {Choi}}, \bibinfo {author}
  {\bibfnamefont {H.}~\bibnamefont {Pichler}},\ and\ \bibinfo {author}
  {\bibfnamefont {M.~D.}\ \bibnamefont {Lukin}},\ }\bibfield  {title} {\bibinfo
  {title} {Quantum approximate optimization algorithm: Performance, mechanism,
  and implementation on near-term devices},\ }\href
  {https://doi.org/10.1103/PhysRevX.10.021067} {\bibfield  {journal} {\bibinfo
  {journal} {Phys. Rev. X}\ }\textbf {\bibinfo {volume} {10}},\ \bibinfo
  {pages} {021067} (\bibinfo {year} {2020})}\BibitemShut {NoStop}%
\bibitem [{\citenamefont {{Crooks}}(2018)}]{crooks2018performance}%
  \BibitemOpen
  \bibfield  {author} {\bibinfo {author} {\bibfnamefont {G.~E.}\ \bibnamefont
  {{Crooks}}},\ }\bibfield  {title} {\bibinfo {title} {{Performance of the
  Quantum Approximate Optimization Algorithm on the Maximum Cut Problem}},\
  }\href@noop {} {\bibfield  {journal} {\bibinfo  {journal} {arXiv e-prints}\
  ,\ \bibinfo {eid} {arXiv:1811.08419}} (\bibinfo {year} {2018})},\ \Eprint
  {https://arxiv.org/abs/1811.08419} {arXiv:1811.08419 [quant-ph]} \BibitemShut
  {NoStop}%
\bibitem [{\citenamefont {{Willsch}}\ \emph {et~al.}(2020)\citenamefont
  {{Willsch}}, \citenamefont {{Willsch}}, \citenamefont {{Jin}}, \citenamefont
  {{De Raedt}},\ and\ \citenamefont {{Michielsen}}}]{willsch2020benchmarking}%
  \BibitemOpen
  \bibfield  {author} {\bibinfo {author} {\bibfnamefont {M.}~\bibnamefont
  {{Willsch}}}, \bibinfo {author} {\bibfnamefont {D.}~\bibnamefont
  {{Willsch}}}, \bibinfo {author} {\bibfnamefont {F.}~\bibnamefont {{Jin}}},
  \bibinfo {author} {\bibfnamefont {H.}~\bibnamefont {{De Raedt}}},\ and\
  \bibinfo {author} {\bibfnamefont {K.}~\bibnamefont {{Michielsen}}},\
  }\bibfield  {title} {\bibinfo {title} {{Benchmarking the quantum approximate
  optimization algorithm}},\ }\href
  {https://doi.org/10.1007/s11128-020-02692-8} {\bibfield  {journal} {\bibinfo
  {journal} {Quantum Information Processing}\ }\textbf {\bibinfo {volume}
  {19}},\ \bibinfo {eid} {197} (\bibinfo {year} {2020})},\ \Eprint
  {https://arxiv.org/abs/1907.02359} {arXiv:1907.02359 [quant-ph]} \BibitemShut
  {NoStop}%
\bibitem [{\citenamefont {{Bravyi}}\ \emph {et~al.}(2019)\citenamefont
  {{Bravyi}}, \citenamefont {{Kliesch}}, \citenamefont {{Koenig}},\ and\
  \citenamefont {{Tang}}}]{bravyi2019obstacles}%
  \BibitemOpen
  \bibfield  {author} {\bibinfo {author} {\bibfnamefont {S.}~\bibnamefont
  {{Bravyi}}}, \bibinfo {author} {\bibfnamefont {A.}~\bibnamefont {{Kliesch}}},
  \bibinfo {author} {\bibfnamefont {R.}~\bibnamefont {{Koenig}}},\ and\
  \bibinfo {author} {\bibfnamefont {E.}~\bibnamefont {{Tang}}},\ }\bibfield
  {title} {\bibinfo {title} {{Obstacles to State Preparation and Variational
  Optimization from Symmetry Protection}},\ }\href@noop {} {\bibfield
  {journal} {\bibinfo  {journal} {arXiv e-prints}\ ,\ \bibinfo {eid}
  {arXiv:1910.08980}} (\bibinfo {year} {2019})},\ \Eprint
  {https://arxiv.org/abs/1910.08980} {arXiv:1910.08980 [quant-ph]} \BibitemShut
  {NoStop}%
\bibitem [{\citenamefont {Guerreschi}\ and\ \citenamefont
  {Matsuura}(2019)}]{guerreschi2019qaoa}%
  \BibitemOpen
  \bibfield  {author} {\bibinfo {author} {\bibfnamefont {G.~G.}\ \bibnamefont
  {Guerreschi}}\ and\ \bibinfo {author} {\bibfnamefont {A.~Y.}\ \bibnamefont
  {Matsuura}},\ }\bibfield  {title} {\bibinfo {title} {{QAOA} for {Max-Cut}
  requires hundreds of qubits for quantum speed-up},\ }\href
  {https://doi.org/10.1038/s41598-019-43176-9} {\bibfield  {journal} {\bibinfo
  {journal} {Scientific Reports}\ }\textbf {\bibinfo {volume} {9}},\ \bibinfo
  {pages} {6903} (\bibinfo {year} {2019})}\BibitemShut {NoStop}%
\bibitem [{\citenamefont {{Shaydulin}}\ \emph {et~al.}(2019)\citenamefont
  {{Shaydulin}}, \citenamefont {{Safro}},\ and\ \citenamefont
  {{Larson}}}]{shaydulin2019multistart}%
  \BibitemOpen
  \bibfield  {author} {\bibinfo {author} {\bibfnamefont {R.}~\bibnamefont
  {{Shaydulin}}}, \bibinfo {author} {\bibfnamefont {I.}~\bibnamefont
  {{Safro}}},\ and\ \bibinfo {author} {\bibfnamefont {J.}~\bibnamefont
  {{Larson}}},\ }\bibfield  {title} {\bibinfo {title} {Multistart methods for
  quantum approximate optimization},\ }in\ \href
  {https://doi.org/10.1109/HPEC.2019.8916288} {\emph {\bibinfo {booktitle}
  {2019 IEEE High Performance Extreme Computing Conference (HPEC)}}}\ (\bibinfo
  {year} {2019})\ pp.\ \bibinfo {pages} {1--8}\BibitemShut {NoStop}%
\bibitem [{\citenamefont {{McClean}}\ \emph {et~al.}(2018)\citenamefont
  {{McClean}}, \citenamefont {{Boixo}}, \citenamefont {{Smelyanskiy}},
  \citenamefont {{Babbush}},\ and\ \citenamefont
  {{Neven}}}]{mcClean2021barren}%
  \BibitemOpen
  \bibfield  {author} {\bibinfo {author} {\bibfnamefont {J.~R.}\ \bibnamefont
  {{McClean}}}, \bibinfo {author} {\bibfnamefont {S.}~\bibnamefont {{Boixo}}},
  \bibinfo {author} {\bibfnamefont {V.~N.}\ \bibnamefont {{Smelyanskiy}}},
  \bibinfo {author} {\bibfnamefont {R.}~\bibnamefont {{Babbush}}},\ and\
  \bibinfo {author} {\bibfnamefont {H.}~\bibnamefont {{Neven}}},\ }\bibfield
  {title} {\bibinfo {title} {{Barren plateaus in quantum neural network
  training landscapes}},\ }\href {https://doi.org/10.1038/s41467-018-07090-4}
  {\bibfield  {journal} {\bibinfo  {journal} {Nature Communications}\ }\textbf
  {\bibinfo {volume} {9}},\ \bibinfo {eid} {4812} (\bibinfo {year} {2018})},\
  \Eprint {https://arxiv.org/abs/1803.11173} {arXiv:1803.11173 [quant-ph]}
  \BibitemShut {NoStop}%
\bibitem [{\citenamefont {{Holmes}}\ \emph {et~al.}(2021)\citenamefont
  {{Holmes}}, \citenamefont {{Sharma}}, \citenamefont {{Cerezo}},\ and\
  \citenamefont {{Coles}}}]{holmes2021connecting}%
  \BibitemOpen
  \bibfield  {author} {\bibinfo {author} {\bibfnamefont {Z.}~\bibnamefont
  {{Holmes}}}, \bibinfo {author} {\bibfnamefont {K.}~\bibnamefont {{Sharma}}},
  \bibinfo {author} {\bibfnamefont {M.}~\bibnamefont {{Cerezo}}},\ and\
  \bibinfo {author} {\bibfnamefont {P.~J.}\ \bibnamefont {{Coles}}},\
  }\bibfield  {title} {\bibinfo {title} {{Connecting ansatz expressibility to
  gradient magnitudes and barren plateaus}},\ }\href@noop {} {\bibfield
  {journal} {\bibinfo  {journal} {arXiv e-prints}\ ,\ \bibinfo {eid}
  {arXiv:2101.02138}} (\bibinfo {year} {2021})},\ \Eprint
  {https://arxiv.org/abs/2101.02138} {arXiv:2101.02138 [quant-ph]} \BibitemShut
  {NoStop}%
\bibitem [{\citenamefont {{Cerezo}}\ \emph {et~al.}(2021)\citenamefont
  {{Cerezo}}, \citenamefont {{Sone}}, \citenamefont {{Volkoff}}, \citenamefont
  {{Cincio}},\ and\ \citenamefont {{Coles}}}]{cerzo2021cost}%
  \BibitemOpen
  \bibfield  {author} {\bibinfo {author} {\bibfnamefont {M.}~\bibnamefont
  {{Cerezo}}}, \bibinfo {author} {\bibfnamefont {A.}~\bibnamefont {{Sone}}},
  \bibinfo {author} {\bibfnamefont {T.}~\bibnamefont {{Volkoff}}}, \bibinfo
  {author} {\bibfnamefont {L.}~\bibnamefont {{Cincio}}},\ and\ \bibinfo
  {author} {\bibfnamefont {P.~J.}\ \bibnamefont {{Coles}}},\ }\bibfield
  {title} {\bibinfo {title} {{Cost function dependent barren plateaus in
  shallow parametrized quantum circuits}},\ }\href
  {https://doi.org/10.1038/s41467-021-21728-w} {\bibfield  {journal} {\bibinfo
  {journal} {Nature Communications}\ }\textbf {\bibinfo {volume} {12}},\
  \bibinfo {eid} {1791} (\bibinfo {year} {2021})},\ \Eprint
  {https://arxiv.org/abs/2001.00550} {arXiv:2001.00550 [quant-ph]} \BibitemShut
  {NoStop}%
\bibitem [{\citenamefont {{Brandao}}\ \emph {et~al.}(2018)\citenamefont
  {{Brandao}}, \citenamefont {{Broughton}}, \citenamefont {{Farhi}},
  \citenamefont {{Gutmann}},\ and\ \citenamefont {{Neven}}}]{Brandao}%
  \BibitemOpen
  \bibfield  {author} {\bibinfo {author} {\bibfnamefont {F.~G.~S.~L.}\
  \bibnamefont {{Brandao}}}, \bibinfo {author} {\bibfnamefont {M.}~\bibnamefont
  {{Broughton}}}, \bibinfo {author} {\bibfnamefont {E.}~\bibnamefont
  {{Farhi}}}, \bibinfo {author} {\bibfnamefont {S.}~\bibnamefont {{Gutmann}}},\
  and\ \bibinfo {author} {\bibfnamefont {H.}~\bibnamefont {{Neven}}},\
  }\bibfield  {title} {\bibinfo {title} {{For Fixed Control Parameters the
  Quantum Approximate Optimization Algorithm's Objective Function Value
  Concentrates for Typical Instances}},\ }\href@noop {} {\bibfield  {journal}
  {\bibinfo  {journal} {arXiv e-prints}\ ,\ \bibinfo {eid} {arXiv:1812.04170}}
  (\bibinfo {year} {2018})},\ \Eprint {https://arxiv.org/abs/1812.04170}
  {arXiv:1812.04170 [quant-ph]} \BibitemShut {NoStop}%
\bibitem [{\citenamefont {{Egger}}\ \emph {et~al.}(2020)\citenamefont
  {{Egger}}, \citenamefont {{Marecek}},\ and\ \citenamefont
  {{Woerner}}}]{egger2020warm}%
  \BibitemOpen
  \bibfield  {author} {\bibinfo {author} {\bibfnamefont {D.~J.}\ \bibnamefont
  {{Egger}}}, \bibinfo {author} {\bibfnamefont {J.}~\bibnamefont {{Marecek}}},\
  and\ \bibinfo {author} {\bibfnamefont {S.}~\bibnamefont {{Woerner}}},\
  }\bibfield  {title} {\bibinfo {title} {{Warm-starting quantum
  optimization}},\ }\href@noop {} {\bibfield  {journal} {\bibinfo  {journal}
  {arXiv e-prints}\ ,\ \bibinfo {eid} {arXiv:2009.10095}} (\bibinfo {year}
  {2020})},\ \Eprint {https://arxiv.org/abs/2009.10095} {arXiv:2009.10095
  [quant-ph]} \BibitemShut {NoStop}%
\bibitem [{\citenamefont {{Alam}}\ \emph {et~al.}(2020)\citenamefont {{Alam}},
  \citenamefont {{Ash-Saki}},\ and\ \citenamefont
  {{Ghosh}}}]{alam2020accelerating}%
  \BibitemOpen
  \bibfield  {author} {\bibinfo {author} {\bibfnamefont {M.}~\bibnamefont
  {{Alam}}}, \bibinfo {author} {\bibfnamefont {A.}~\bibnamefont {{Ash-Saki}}},\
  and\ \bibinfo {author} {\bibfnamefont {S.}~\bibnamefont {{Ghosh}}},\
  }\bibfield  {title} {\bibinfo {title} {{Accelerating Quantum Approximate
  Optimization Algorithm using Machine Learning}},\ }\href@noop {} {\bibfield
  {journal} {\bibinfo  {journal} {arXiv e-prints}\ ,\ \bibinfo {eid}
  {arXiv:2002.01089}} (\bibinfo {year} {2020})},\ \Eprint
  {https://arxiv.org/abs/2002.01089} {arXiv:2002.01089 [cs.ET]} \BibitemShut
  {NoStop}%
\bibitem [{\citenamefont {{Khairy}}\ \emph {et~al.}(2019)\citenamefont
  {{Khairy}}, \citenamefont {{Shaydulin}}, \citenamefont {{Cincio}},
  \citenamefont {{Alexeev}},\ and\ \citenamefont
  {{Balaprakash}}}]{khairy2019learning}%
  \BibitemOpen
  \bibfield  {author} {\bibinfo {author} {\bibfnamefont {S.}~\bibnamefont
  {{Khairy}}}, \bibinfo {author} {\bibfnamefont {R.}~\bibnamefont
  {{Shaydulin}}}, \bibinfo {author} {\bibfnamefont {L.}~\bibnamefont
  {{Cincio}}}, \bibinfo {author} {\bibfnamefont {Y.}~\bibnamefont
  {{Alexeev}}},\ and\ \bibinfo {author} {\bibfnamefont {P.}~\bibnamefont
  {{Balaprakash}}},\ }\bibfield  {title} {\bibinfo {title} {{Learning to
  Optimize Variational Quantum Circuits to Solve Combinatorial Problems}},\
  }\href@noop {} {\bibfield  {journal} {\bibinfo  {journal} {arXiv e-prints}\
  ,\ \bibinfo {eid} {arXiv:1911.11071}} (\bibinfo {year} {2019})},\ \Eprint
  {https://arxiv.org/abs/1911.11071} {arXiv:1911.11071 [cs.LG]} \BibitemShut
  {NoStop}%
\bibitem [{\citenamefont {Albash}\ and\ \citenamefont
  {Lidar}(2018)}]{albash2018adiabatic}%
  \BibitemOpen
  \bibfield  {author} {\bibinfo {author} {\bibfnamefont {T.}~\bibnamefont
  {Albash}}\ and\ \bibinfo {author} {\bibfnamefont {D.~A.}\ \bibnamefont
  {Lidar}},\ }\bibfield  {title} {\bibinfo {title} {Adiabatic quantum
  computation},\ }\href {https://doi.org/10.1103/RevModPhys.90.015002}
  {\bibfield  {journal} {\bibinfo  {journal} {Rev. Mod. Phys.}\ }\textbf
  {\bibinfo {volume} {90}},\ \bibinfo {pages} {015002} (\bibinfo {year}
  {2018})}\BibitemShut {NoStop}%
\bibitem [{\citenamefont {Liang}\ \emph {et~al.}(2020)\citenamefont {Liang},
  \citenamefont {Li},\ and\ \citenamefont
  {Leichenauer}}]{liang2020investigating}%
  \BibitemOpen
  \bibfield  {author} {\bibinfo {author} {\bibfnamefont {D.}~\bibnamefont
  {Liang}}, \bibinfo {author} {\bibfnamefont {L.}~\bibnamefont {Li}},\ and\
  \bibinfo {author} {\bibfnamefont {S.}~\bibnamefont {Leichenauer}},\
  }\bibfield  {title} {\bibinfo {title} {Investigating quantum approximate
  optimization algorithms under bang-bang protocols},\ }\href
  {https://doi.org/10.1103/PhysRevResearch.2.033402} {\bibfield  {journal}
  {\bibinfo  {journal} {Phys. Rev. Research}\ }\textbf {\bibinfo {volume}
  {2}},\ \bibinfo {pages} {033402} (\bibinfo {year} {2020})}\BibitemShut
  {NoStop}%
\bibitem [{\citenamefont {Heyl}\ \emph {et~al.}(2019)\citenamefont {Heyl},
  \citenamefont {Hauke},\ and\ \citenamefont {Zoller}}]{heyl2019quantum}%
  \BibitemOpen
  \bibfield  {author} {\bibinfo {author} {\bibfnamefont {M.}~\bibnamefont
  {Heyl}}, \bibinfo {author} {\bibfnamefont {P.}~\bibnamefont {Hauke}},\ and\
  \bibinfo {author} {\bibfnamefont {P.}~\bibnamefont {Zoller}},\ }\bibfield
  {title} {\bibinfo {title} {Quantum localization bounds Trotter errors in
  digital quantum simulation},\ }\bibfield  {journal} {\bibinfo  {journal}
  {Science Advances}\ }\textbf {\bibinfo {volume} {5}},\ \href
  {https://doi.org/10.1126/sciadv.aau8342} {10.1126/sciadv.aau8342} (\bibinfo
  {year} {2019})\BibitemShut {NoStop}%
\bibitem [{\citenamefont {Goemans}\ and\ \citenamefont
  {Williamson}(1995)}]{goemans1995improved}%
  \BibitemOpen
  \bibfield  {author} {\bibinfo {author} {\bibfnamefont {M.~X.}\ \bibnamefont
  {Goemans}}\ and\ \bibinfo {author} {\bibfnamefont {D.~P.}\ \bibnamefont
  {Williamson}},\ }\bibfield  {title} {\bibinfo {title} {Improved approximation
  algorithms for maximum cut and satisfiability problems using semidefinite
  programming},\ }\href {https://doi.org/10.1145/227683.227684} {\bibfield
  {journal} {\bibinfo  {journal} {J. ACM}\ }\textbf {\bibinfo {volume} {42}},\
  \bibinfo {pages} {1115} (\bibinfo {year} {1995})}\BibitemShut {NoStop}%
\bibitem [{\citenamefont {{Wurtz}}\ and\ \citenamefont
  {{Love}}(2020)}]{wurtz2020bounds}%
  \BibitemOpen
  \bibfield  {author} {\bibinfo {author} {\bibfnamefont {J.}~\bibnamefont
  {{Wurtz}}}\ and\ \bibinfo {author} {\bibfnamefont {P.~J.}\ \bibnamefont
  {{Love}}},\ }\bibfield  {title} {\bibinfo {title} {{Bounds on MAXCUT QAOA
  performance for $p>1$}},\ }\href@noop {} {\bibfield  {journal} {\bibinfo
  {journal} {arXiv e-prints}\ ,\ \bibinfo {eid} {arXiv:2010.11209}} (\bibinfo
  {year} {2020})},\ \Eprint {https://arxiv.org/abs/2010.11209}
  {arXiv:2010.11209 [quant-ph]} \BibitemShut {NoStop}%
\bibitem [{\citenamefont {BROYDEN}(1970)}]{bfgs1}%
  \BibitemOpen
  \bibfield  {author} {\bibinfo {author} {\bibfnamefont {C.~G.}\ \bibnamefont
  {BROYDEN}},\ }\bibfield  {title} {\bibinfo {title} {{The Convergence of a
  Class of Double-rank Minimization Algorithms 1. General Considerations}},\
  }\href {https://doi.org/10.1093/imamat/6.1.76} {\bibfield  {journal}
  {\bibinfo  {journal} {IMA Journal of Applied Mathematics}\ }\textbf {\bibinfo
  {volume} {6}},\ \bibinfo {pages} {76} (\bibinfo {year} {1970})}\BibitemShut
  {NoStop}%
\bibitem [{\citenamefont {Fletcher}(1970)}]{bfgs2}%
  \BibitemOpen
  \bibfield  {author} {\bibinfo {author} {\bibfnamefont {R.}~\bibnamefont
  {Fletcher}},\ }\bibfield  {title} {\bibinfo {title} {{A new approach to
  variable metric algorithms}},\ }\href
  {https://doi.org/10.1093/comjnl/13.3.317} {\bibfield  {journal} {\bibinfo
  {journal} {The Computer Journal}\ }\textbf {\bibinfo {volume} {13}},\
  \bibinfo {pages} {317} (\bibinfo {year} {1970})}\BibitemShut {NoStop}%
\bibitem [{\citenamefont {Goldfarb}(1970)}]{bfgs3}%
  \BibitemOpen
  \bibfield  {author} {\bibinfo {author} {\bibfnamefont {D.}~\bibnamefont
  {Goldfarb}},\ }\bibfield  {title} {\bibinfo {title} {A family of
  variable-metric methods derived by variational means},\ }\href
  {http://www.jstor.org/stable/2004873} {\bibfield  {journal} {\bibinfo
  {journal} {Mathematics of Computation}\ }\textbf {\bibinfo {volume} {24}},\
  \bibinfo {pages} {23} (\bibinfo {year} {1970})}\BibitemShut {NoStop}%
\bibitem [{\citenamefont {Shanno}(1970)}]{bfgs4}%
  \BibitemOpen
  \bibfield  {author} {\bibinfo {author} {\bibfnamefont {D.~F.}\ \bibnamefont
  {Shanno}},\ }\bibfield  {title} {\bibinfo {title} {Conditioning of
  quasi-Newton methods for function minimization},\ }\href
  {http://www.jstor.org/stable/2004840} {\bibfield  {journal} {\bibinfo
  {journal} {Mathematics of Computation}\ }\textbf {\bibinfo {volume} {24}},\
  \bibinfo {pages} {647} (\bibinfo {year} {1970})}\BibitemShut {NoStop}%
\bibitem [{\citenamefont {Virtanen}\ \emph {et~al.}(2020)\citenamefont
  {Virtanen}, \citenamefont {Gommers}, \citenamefont {Oliphant}, \citenamefont
  {Haberland}, \citenamefont {Reddy}, \citenamefont {Cournapeau}, \citenamefont
  {Burovski}, \citenamefont {Peterson}, \citenamefont {Weckesser},
  \citenamefont {Bright}, \citenamefont {{van der Walt}}, \citenamefont
  {Brett}, \citenamefont {Wilson}, \citenamefont {Millman}, \citenamefont
  {Mayorov}, \citenamefont {Nelson}, \citenamefont {Jones}, \citenamefont
  {Kern}, \citenamefont {Larson}, \citenamefont {Carey}, \citenamefont {Polat},
  \citenamefont {Feng}, \citenamefont {Moore}, \citenamefont {{VanderPlas}},
  \citenamefont {Laxalde}, \citenamefont {Perktold}, \citenamefont {Cimrman},
  \citenamefont {Henriksen}, \citenamefont {Quintero}, \citenamefont {Harris},
  \citenamefont {Archibald}, \citenamefont {Ribeiro}, \citenamefont
  {Pedregosa}, \citenamefont {{van Mulbregt}},\ and\ \citenamefont {{SciPy 1.0
  Contributors}}}]{scipy}%
  \BibitemOpen
  \bibfield  {author} {\bibinfo {author} {\bibfnamefont {P.}~\bibnamefont
  {Virtanen}}, \bibinfo {author} {\bibfnamefont {R.}~\bibnamefont {Gommers}},
  \bibinfo {author} {\bibfnamefont {T.~E.}\ \bibnamefont {Oliphant}}, \bibinfo
  {author} {\bibfnamefont {M.}~\bibnamefont {Haberland}}, \bibinfo {author}
  {\bibfnamefont {T.}~\bibnamefont {Reddy}}, \bibinfo {author} {\bibfnamefont
  {D.}~\bibnamefont {Cournapeau}}, \bibinfo {author} {\bibfnamefont
  {E.}~\bibnamefont {Burovski}}, \bibinfo {author} {\bibfnamefont
  {P.}~\bibnamefont {Peterson}}, \bibinfo {author} {\bibfnamefont
  {W.}~\bibnamefont {Weckesser}}, \bibinfo {author} {\bibfnamefont
  {J.}~\bibnamefont {Bright}}, \bibinfo {author} {\bibfnamefont {S.~J.}\
  \bibnamefont {{van der Walt}}}, \bibinfo {author} {\bibfnamefont
  {M.}~\bibnamefont {Brett}}, \bibinfo {author} {\bibfnamefont
  {J.}~\bibnamefont {Wilson}}, \bibinfo {author} {\bibfnamefont {K.~J.}\
  \bibnamefont {Millman}}, \bibinfo {author} {\bibfnamefont {N.}~\bibnamefont
  {Mayorov}}, \bibinfo {author} {\bibfnamefont {A.~R.~J.}\ \bibnamefont
  {Nelson}}, \bibinfo {author} {\bibfnamefont {E.}~\bibnamefont {Jones}},
  \bibinfo {author} {\bibfnamefont {R.}~\bibnamefont {Kern}}, \bibinfo {author}
  {\bibfnamefont {E.}~\bibnamefont {Larson}}, \bibinfo {author} {\bibfnamefont
  {C.~J.}\ \bibnamefont {Carey}}, \bibinfo {author} {\bibfnamefont
  {{\.I}.}~\bibnamefont {Polat}}, \bibinfo {author} {\bibfnamefont
  {Y.}~\bibnamefont {Feng}}, \bibinfo {author} {\bibfnamefont {E.~W.}\
  \bibnamefont {Moore}}, \bibinfo {author} {\bibfnamefont {J.}~\bibnamefont
  {{VanderPlas}}}, \bibinfo {author} {\bibfnamefont {D.}~\bibnamefont
  {Laxalde}}, \bibinfo {author} {\bibfnamefont {J.}~\bibnamefont {Perktold}},
  \bibinfo {author} {\bibfnamefont {R.}~\bibnamefont {Cimrman}}, \bibinfo
  {author} {\bibfnamefont {I.}~\bibnamefont {Henriksen}}, \bibinfo {author}
  {\bibfnamefont {E.~A.}\ \bibnamefont {Quintero}}, \bibinfo {author}
  {\bibfnamefont {C.~R.}\ \bibnamefont {Harris}}, \bibinfo {author}
  {\bibfnamefont {A.~M.}\ \bibnamefont {Archibald}}, \bibinfo {author}
  {\bibfnamefont {A.~H.}\ \bibnamefont {Ribeiro}}, \bibinfo {author}
  {\bibfnamefont {F.}~\bibnamefont {Pedregosa}}, \bibinfo {author}
  {\bibfnamefont {P.}~\bibnamefont {{van Mulbregt}}},\ and\ \bibinfo {author}
  {\bibnamefont {{SciPy 1.0 Contributors}}},\ }\bibfield  {title} {\bibinfo
  {title} {{{SciPy} 1.0: Fundamental Algorithms for Scientific Computing in
  Python}},\ }\href {https://doi.org/10.1038/s41592-019-0686-2} {\bibfield
  {journal} {\bibinfo  {journal} {Nature Methods}\ }\textbf {\bibinfo {volume}
  {17}},\ \bibinfo {pages} {261} (\bibinfo {year} {2020})}\BibitemShut
  {NoStop}%
\bibitem [{\citenamefont {Rosenblatt}(1956)}]{rosenblatt1956}%
  \BibitemOpen
  \bibfield  {author} {\bibinfo {author} {\bibfnamefont {M.}~\bibnamefont
  {Rosenblatt}},\ }\bibfield  {title} {\bibinfo {title} {Remarks on some
  nonparametric estimates of a density function},\ }\href
  {https://doi.org/10.1214/aoms/1177728190} {\bibfield  {journal} {\bibinfo
  {journal} {Ann. Math. Statist.}\ }\textbf {\bibinfo {volume} {27}},\ \bibinfo
  {pages} {832} (\bibinfo {year} {1956})}\BibitemShut {NoStop}%
\bibitem [{\citenamefont {Parzen}(1962)}]{parzen1962}%
  \BibitemOpen
  \bibfield  {author} {\bibinfo {author} {\bibfnamefont {E.}~\bibnamefont
  {Parzen}},\ }\bibfield  {title} {\bibinfo {title} {On estimation of a
  probability density function and mode},\ }\href
  {https://doi.org/10.1214/aoms/1177704472} {\bibfield  {journal} {\bibinfo
  {journal} {Ann. Math. Statist.}\ }\textbf {\bibinfo {volume} {33}},\ \bibinfo
  {pages} {1065} (\bibinfo {year} {1962})}\BibitemShut {NoStop}%
\bibitem [{\citenamefont {Kadowaki}\ and\ \citenamefont
  {Nishimori}(1998)}]{kadowaki1998quantum}%
  \BibitemOpen
  \bibfield  {author} {\bibinfo {author} {\bibfnamefont {T.}~\bibnamefont
  {Kadowaki}}\ and\ \bibinfo {author} {\bibfnamefont {H.}~\bibnamefont
  {Nishimori}},\ }\bibfield  {title} {\bibinfo {title} {Quantum annealing in
  the transverse Ising model},\ }\href
  {https://doi.org/10.1103/PhysRevE.58.5355} {\bibfield  {journal} {\bibinfo
  {journal} {Phys. Rev. E}\ }\textbf {\bibinfo {volume} {58}},\ \bibinfo
  {pages} {5355} (\bibinfo {year} {1998})}\BibitemShut {NoStop}%
\bibitem [{\citenamefont {{Brooke}}\ \emph {et~al.}(1999)\citenamefont
  {{Brooke}}, \citenamefont {{Bitko}}, \citenamefont {{Rosenbaum}},\ and\
  \citenamefont {{Aeppli}}}]{brooke1999quantum}%
  \BibitemOpen
  \bibfield  {author} {\bibinfo {author} {\bibfnamefont {J.}~\bibnamefont
  {{Brooke}}}, \bibinfo {author} {\bibfnamefont {D.}~\bibnamefont {{Bitko}}},
  \bibinfo {author} {\bibfnamefont {T.~F.}\ \bibnamefont {{Rosenbaum}}},\ and\
  \bibinfo {author} {\bibfnamefont {G.}~\bibnamefont {{Aeppli}}},\ }\bibfield
  {title} {\bibinfo {title} {{Quantum Annealing of a Disordered Magnet}},\
  }\href {https://doi.org/10.1126/science.284.5415.779} {\bibfield  {journal}
  {\bibinfo  {journal} {Science}\ }\textbf {\bibinfo {volume} {284}},\ \bibinfo
  {pages} {779} (\bibinfo {year} {1999})}\BibitemShut {NoStop}%
\bibitem [{\citenamefont {{Farhi}}\ \emph {et~al.}(2001)\citenamefont
  {{Farhi}}, \citenamefont {{Goldstone}}, \citenamefont {{Gutmann}},
  \citenamefont {{Lapan}}, \citenamefont {{Lundgren}},\ and\ \citenamefont
  {{Preda}}}]{farhi2001quantum}%
  \BibitemOpen
  \bibfield  {author} {\bibinfo {author} {\bibfnamefont {E.}~\bibnamefont
  {{Farhi}}}, \bibinfo {author} {\bibfnamefont {J.}~\bibnamefont
  {{Goldstone}}}, \bibinfo {author} {\bibfnamefont {S.}~\bibnamefont
  {{Gutmann}}}, \bibinfo {author} {\bibfnamefont {J.}~\bibnamefont {{Lapan}}},
  \bibinfo {author} {\bibfnamefont {A.}~\bibnamefont {{Lundgren}}},\ and\
  \bibinfo {author} {\bibfnamefont {D.}~\bibnamefont {{Preda}}},\ }\bibfield
  {title} {\bibinfo {title} {{A Quantum Adiabatic Evolution Algorithm Applied
  to Random Instances of an NP-Complete Problem}},\ }\href
  {https://doi.org/10.1126/science.1057726} {\bibfield  {journal} {\bibinfo
  {journal} {Science}\ }\textbf {\bibinfo {volume} {292}},\ \bibinfo {pages}
  {472} (\bibinfo {year} {2001})},\ \Eprint
  {https://arxiv.org/abs/quant-ph/0104129} {arXiv:quant-ph/0104129 [quant-ph]}
  \BibitemShut {NoStop}%
\bibitem [{\citenamefont {{Farhi}}\ \emph {et~al.}(2000)\citenamefont
  {{Farhi}}, \citenamefont {{Goldstone}}, \citenamefont {{Gutmann}},\ and\
  \citenamefont {{Sipser}}}]{farhi2000quantum}%
  \BibitemOpen
  \bibfield  {author} {\bibinfo {author} {\bibfnamefont {E.}~\bibnamefont
  {{Farhi}}}, \bibinfo {author} {\bibfnamefont {J.}~\bibnamefont
  {{Goldstone}}}, \bibinfo {author} {\bibfnamefont {S.}~\bibnamefont
  {{Gutmann}}},\ and\ \bibinfo {author} {\bibfnamefont {M.}~\bibnamefont
  {{Sipser}}},\ }\bibfield  {title} {\bibinfo {title} {{Quantum Computation by
  Adiabatic Evolution}},\ }\href@noop {} {\bibfield  {journal} {\bibinfo
  {journal} {arXiv e-prints}\ ,\ \bibinfo {eid} {quant-ph/0001106}} (\bibinfo
  {year} {2000})},\ \Eprint {https://arxiv.org/abs/quant-ph/0001106}
  {arXiv:quant-ph/0001106 [quant-ph]} \BibitemShut {NoStop}%
\bibitem [{\citenamefont {{Aharonov}}\ \emph {et~al.}(2008)\citenamefont
  {{Aharonov}}, \citenamefont {{van Dam}}, \citenamefont {{Kempe}},
  \citenamefont {{Landau}}, \citenamefont {{Lloyd}},\ and\ \citenamefont
  {{Regev}}}]{aharonov2008adiabatic}%
  \BibitemOpen
  \bibfield  {author} {\bibinfo {author} {\bibfnamefont {D.}~\bibnamefont
  {{Aharonov}}}, \bibinfo {author} {\bibfnamefont {W.}~\bibnamefont {{van
  Dam}}}, \bibinfo {author} {\bibfnamefont {J.}~\bibnamefont {{Kempe}}},
  \bibinfo {author} {\bibfnamefont {Z.}~\bibnamefont {{Landau}}}, \bibinfo
  {author} {\bibfnamefont {S.}~\bibnamefont {{Lloyd}}},\ and\ \bibinfo {author}
  {\bibfnamefont {O.}~\bibnamefont {{Regev}}},\ }\bibfield  {title} {\bibinfo
  {title} {{Adiabatic Quantum Computation Is Equivalent to Standard Quantum
  Computation}},\ }\href {https://doi.org/10.1137/080734479} {\bibfield
  {journal} {\bibinfo  {journal} {SIAM Review}\ }\textbf {\bibinfo {volume}
  {50}},\ \bibinfo {pages} {755} (\bibinfo {year} {2008})}\BibitemShut
  {NoStop}%
\bibitem [{\citenamefont {{Smith}}\ \emph {et~al.}(2019)\citenamefont
  {{Smith}}, \citenamefont {{Kim}}, \citenamefont {{Pollmann}},\ and\
  \citenamefont {{Knolle}}}]{smith2019simulating}%
  \BibitemOpen
  \bibfield  {author} {\bibinfo {author} {\bibfnamefont {A.}~\bibnamefont
  {{Smith}}}, \bibinfo {author} {\bibfnamefont {M.~S.}\ \bibnamefont {{Kim}}},
  \bibinfo {author} {\bibfnamefont {F.}~\bibnamefont {{Pollmann}}},\ and\
  \bibinfo {author} {\bibfnamefont {J.}~\bibnamefont {{Knolle}}},\ }\bibfield
  {title} {\bibinfo {title} {{Simulating quantum many-body dynamics on a
  current digital quantum computer}},\ }\href
  {https://doi.org/10.1038/s41534-019-0217-0} {\bibfield  {journal} {\bibinfo
  {journal} {npj Quantum Information}\ }\textbf {\bibinfo {volume} {5}},\
  \bibinfo {eid} {106} (\bibinfo {year} {2019})},\ \Eprint
  {https://arxiv.org/abs/1906.06343} {arXiv:1906.06343 [quant-ph]} \BibitemShut
  {NoStop}%
\bibitem [{\citenamefont {{Schollw{\"o}ck}}(2011)}]{schollwock2011density}%
  \BibitemOpen
  \bibfield  {author} {\bibinfo {author} {\bibfnamefont {U.}~\bibnamefont
  {{Schollw{\"o}ck}}},\ }\bibfield  {title} {\bibinfo {title} {{The
  density-matrix renormalization group in the age of matrix product states}},\
  }\href {https://doi.org/10.1016/j.aop.2010.09.012} {\bibfield  {journal}
  {\bibinfo  {journal} {Annals of Physics}\ }\textbf {\bibinfo {volume}
  {326}},\ \bibinfo {pages} {96} (\bibinfo {year} {2011})},\ \Eprint
  {https://arxiv.org/abs/1008.3477} {arXiv:1008.3477 [cond-mat.str-el]}
  \BibitemShut {NoStop}%
\bibitem [{\citenamefont {{Carleo}}\ and\ \citenamefont
  {{Troyer}}(2017)}]{carleo2017solving}%
  \BibitemOpen
  \bibfield  {author} {\bibinfo {author} {\bibfnamefont {G.}~\bibnamefont
  {{Carleo}}}\ and\ \bibinfo {author} {\bibfnamefont {M.}~\bibnamefont
  {{Troyer}}},\ }\bibfield  {title} {\bibinfo {title} {{Solving the quantum
  many-body problem with artificial neural networks}},\ }\href
  {https://doi.org/10.1126/science.aag2302} {\bibfield  {journal} {\bibinfo
  {journal} {Science}\ }\textbf {\bibinfo {volume} {355}},\ \bibinfo {pages}
  {602} (\bibinfo {year} {2017})},\ \Eprint {https://arxiv.org/abs/1606.02318}
  {arXiv:1606.02318 [cond-mat.dis-nn]} \BibitemShut {NoStop}%
\bibitem [{\citenamefont {{Medvidovic}}\ and\ \citenamefont
  {{Carleo}}(2020)}]{medvidovic2020classical}%
  \BibitemOpen
  \bibfield  {author} {\bibinfo {author} {\bibfnamefont {M.}~\bibnamefont
  {{Medvidovic}}}\ and\ \bibinfo {author} {\bibfnamefont {G.}~\bibnamefont
  {{Carleo}}},\ }\bibfield  {title} {\bibinfo {title} {{Classical variational
  simulation of the Quantum Approximate Optimization Algorithm}},\ }\href@noop
  {} {\bibfield  {journal} {\bibinfo  {journal} {arXiv e-prints}\ ,\ \bibinfo
  {eid} {arXiv:2009.01760}} (\bibinfo {year} {2020})},\ \Eprint
  {https://arxiv.org/abs/2009.01760} {arXiv:2009.01760 [quant-ph]} \BibitemShut
  {NoStop}%
\bibitem [{\citenamefont {{Abbas}}\ \emph {et~al.}(2020)\citenamefont
  {{Abbas}}, \citenamefont {{Sutter}}, \citenamefont {{Zoufal}}, \citenamefont
  {{Lucchi}}, \citenamefont {{Figalli}},\ and\ \citenamefont
  {{Woerner}}}]{abbas2020power}%
  \BibitemOpen
  \bibfield  {author} {\bibinfo {author} {\bibfnamefont {A.}~\bibnamefont
  {{Abbas}}}, \bibinfo {author} {\bibfnamefont {D.}~\bibnamefont {{Sutter}}},
  \bibinfo {author} {\bibfnamefont {C.}~\bibnamefont {{Zoufal}}}, \bibinfo
  {author} {\bibfnamefont {A.}~\bibnamefont {{Lucchi}}}, \bibinfo {author}
  {\bibfnamefont {A.}~\bibnamefont {{Figalli}}},\ and\ \bibinfo {author}
  {\bibfnamefont {S.}~\bibnamefont {{Woerner}}},\ }\bibfield  {title} {\bibinfo
  {title} {{The power of quantum neural networks}},\ }\href@noop {} {\bibfield
  {journal} {\bibinfo  {journal} {arXiv e-prints}\ ,\ \bibinfo {eid}
  {arXiv:2011.00027}} (\bibinfo {year} {2020})},\ \Eprint
  {https://arxiv.org/abs/2011.00027} {arXiv:2011.00027 [quant-ph]} \BibitemShut
  {NoStop}%
\bibitem [{\citenamefont {Gu\'ery-Odelin}\ \emph {et~al.}(2019)\citenamefont
  {Gu\'ery-Odelin}, \citenamefont {Ruschhaupt}, \citenamefont {Kiely},
  \citenamefont {Torrontegui}, \citenamefont {Mart\'{\i}nez-Garaot},\ and\
  \citenamefont {Muga}}]{STA}%
  \BibitemOpen
  \bibfield  {author} {\bibinfo {author} {\bibfnamefont {D.}~\bibnamefont
  {Gu\'ery-Odelin}}, \bibinfo {author} {\bibfnamefont {A.}~\bibnamefont
  {Ruschhaupt}}, \bibinfo {author} {\bibfnamefont {A.}~\bibnamefont {Kiely}},
  \bibinfo {author} {\bibfnamefont {E.}~\bibnamefont {Torrontegui}}, \bibinfo
  {author} {\bibfnamefont {S.}~\bibnamefont {Mart\'{\i}nez-Garaot}},\ and\
  \bibinfo {author} {\bibfnamefont {J.~G.}\ \bibnamefont {Muga}},\ }\bibfield
  {title} {\bibinfo {title} {Shortcuts to adiabaticity: Concepts, methods, and
  applications},\ }\href {https://doi.org/10.1103/RevModPhys.91.045001}
  {\bibfield  {journal} {\bibinfo  {journal} {Rev. Mod. Phys.}\ }\textbf
  {\bibinfo {volume} {91}},\ \bibinfo {pages} {045001} (\bibinfo {year}
  {2019})}\BibitemShut {NoStop}%
\bibitem [{\citenamefont {Sels}\ and\ \citenamefont
  {Polkovnikov}(2017)}]{Sels}%
  \BibitemOpen
  \bibfield  {author} {\bibinfo {author} {\bibfnamefont {D.}~\bibnamefont
  {Sels}}\ and\ \bibinfo {author} {\bibfnamefont {A.}~\bibnamefont
  {Polkovnikov}},\ }\bibfield  {title} {\bibinfo {title} {Minimizing
  irreversible losses in quantum systems by local counterdiabatic driving},\
  }\href {https://doi.org/10.1073/pnas.1619826114} {\bibfield  {journal}
  {\bibinfo  {journal} {Proceedings of the National Academy of Sciences}\
  }\textbf {\bibinfo {volume} {114}},\ \bibinfo {pages} {E3909} (\bibinfo
  {year} {2017})}\BibitemShut {NoStop}%
\bibitem [{\citenamefont {Claeys}\ \emph {et~al.}(2019)\citenamefont {Claeys},
  \citenamefont {Pandey}, \citenamefont {Sels},\ and\ \citenamefont
  {Polkovnikov}}]{Clayes}%
  \BibitemOpen
  \bibfield  {author} {\bibinfo {author} {\bibfnamefont {P.~W.}\ \bibnamefont
  {Claeys}}, \bibinfo {author} {\bibfnamefont {M.}~\bibnamefont {Pandey}},
  \bibinfo {author} {\bibfnamefont {D.}~\bibnamefont {Sels}},\ and\ \bibinfo
  {author} {\bibfnamefont {A.}~\bibnamefont {Polkovnikov}},\ }\bibfield
  {title} {\bibinfo {title} {Floquet-engineering counterdiabatic protocols in
  quantum many-body systems},\ }\href
  {https://doi.org/10.1103/PhysRevLett.123.090602} {\bibfield  {journal}
  {\bibinfo  {journal} {Phys. Rev. Lett.}\ }\textbf {\bibinfo {volume} {123}},\
  \bibinfo {pages} {090602} (\bibinfo {year} {2019})}\BibitemShut {NoStop}%
\bibitem [{\citenamefont {Yang}\ \emph {et~al.}(2017)\citenamefont {Yang},
  \citenamefont {Rahmani}, \citenamefont {Shabani}, \citenamefont {Neven},\
  and\ \citenamefont {Chamon}}]{Chamon}%
  \BibitemOpen
  \bibfield  {author} {\bibinfo {author} {\bibfnamefont {Z.-C.}\ \bibnamefont
  {Yang}}, \bibinfo {author} {\bibfnamefont {A.}~\bibnamefont {Rahmani}},
  \bibinfo {author} {\bibfnamefont {A.}~\bibnamefont {Shabani}}, \bibinfo
  {author} {\bibfnamefont {H.}~\bibnamefont {Neven}},\ and\ \bibinfo {author}
  {\bibfnamefont {C.}~\bibnamefont {Chamon}},\ }\bibfield  {title} {\bibinfo
  {title} {Optimizing variational quantum algorithms using Pontryagin's minimum
  principle},\ }\href {https://doi.org/10.1103/PhysRevX.7.021027} {\bibfield
  {journal} {\bibinfo  {journal} {Phys. Rev. X}\ }\textbf {\bibinfo {volume}
  {7}},\ \bibinfo {pages} {021027} (\bibinfo {year} {2017})}\BibitemShut
  {NoStop}%
\bibitem [{\citenamefont {Sack}(2021)}]{stefan}%
  \BibitemOpen
  \bibfield  {author} {\bibinfo {author} {\bibfnamefont {S.~H.}\ \bibnamefont
  {Sack}},\ }\href@noop {} {\bibinfo {title} {Trotterized quantum annealing
  initialization of the QAOA}},\ \bibinfo {howpublished}
  {\url{https://github.com/shsack/TQA-init.-for-QAOA}} (\bibinfo {year}
  {2021})\BibitemShut {NoStop}%
\bibitem [{\citenamefont {Hagberg}\ \emph {et~al.}(2008)\citenamefont
  {Hagberg}, \citenamefont {Swart},\ and\ \citenamefont {S~Chult}}]{networkx}%
  \BibitemOpen
  \bibfield  {author} {\bibinfo {author} {\bibfnamefont {A.}~\bibnamefont
  {Hagberg}}, \bibinfo {author} {\bibfnamefont {P.}~\bibnamefont {Swart}},\
  and\ \bibinfo {author} {\bibfnamefont {D.}~\bibnamefont {S~Chult}},\ }\href
  {https://networkx.org/} {\emph {\bibinfo {title} {Exploring network
  structure, dynamics, and function using NetworkX}}},\ \bibinfo {type} {Tech.
  Rep.}\ (\bibinfo  {institution} {Los Alamos National Lab.(LANL), Los Alamos,
  NM (United States)},\ \bibinfo {year} {2008})\BibitemShut {NoStop}%
\bibitem [{\citenamefont {{Berry}}\ \emph {et~al.}(2007)\citenamefont
  {{Berry}}, \citenamefont {{Ahokas}}, \citenamefont {{Cleve}},\ and\
  \citenamefont {{Sanders}}}]{berry2007efficient}%
  \BibitemOpen
  \bibfield  {author} {\bibinfo {author} {\bibfnamefont {D.~W.}\ \bibnamefont
  {{Berry}}}, \bibinfo {author} {\bibfnamefont {G.}~\bibnamefont {{Ahokas}}},
  \bibinfo {author} {\bibfnamefont {R.}~\bibnamefont {{Cleve}}},\ and\ \bibinfo
  {author} {\bibfnamefont {B.~C.}\ \bibnamefont {{Sanders}}},\ }\bibfield
  {title} {\bibinfo {title} {{Efficient Quantum Algorithms for Simulating
  Sparse Hamiltonians}},\ }\href {https://doi.org/10.1007/s00220-006-0150-x}
  {\bibfield  {journal} {\bibinfo  {journal} {Communications in Mathematical
  Physics}\ }\textbf {\bibinfo {volume} {270}},\ \bibinfo {pages} {359}
  (\bibinfo {year} {2007})},\ \Eprint {https://arxiv.org/abs/quant-ph/0508139}
  {arXiv:quant-ph/0508139 [quant-ph]} \BibitemShut {NoStop}%
\end{thebibliography}
\end{document}